Hokkaido University Group Research Achievement Report

# Research and Development of High-Efficiency Compressed Transmission Technologies for Holographic Communication in Beyond 5G Networks

**Authors:** Yuji Sakamoto, Takahiro Yamanoi, and Seok Kang(Hokkaido University)
**Funding Agency:** National Institute of Information and Communications Technology (NICT)

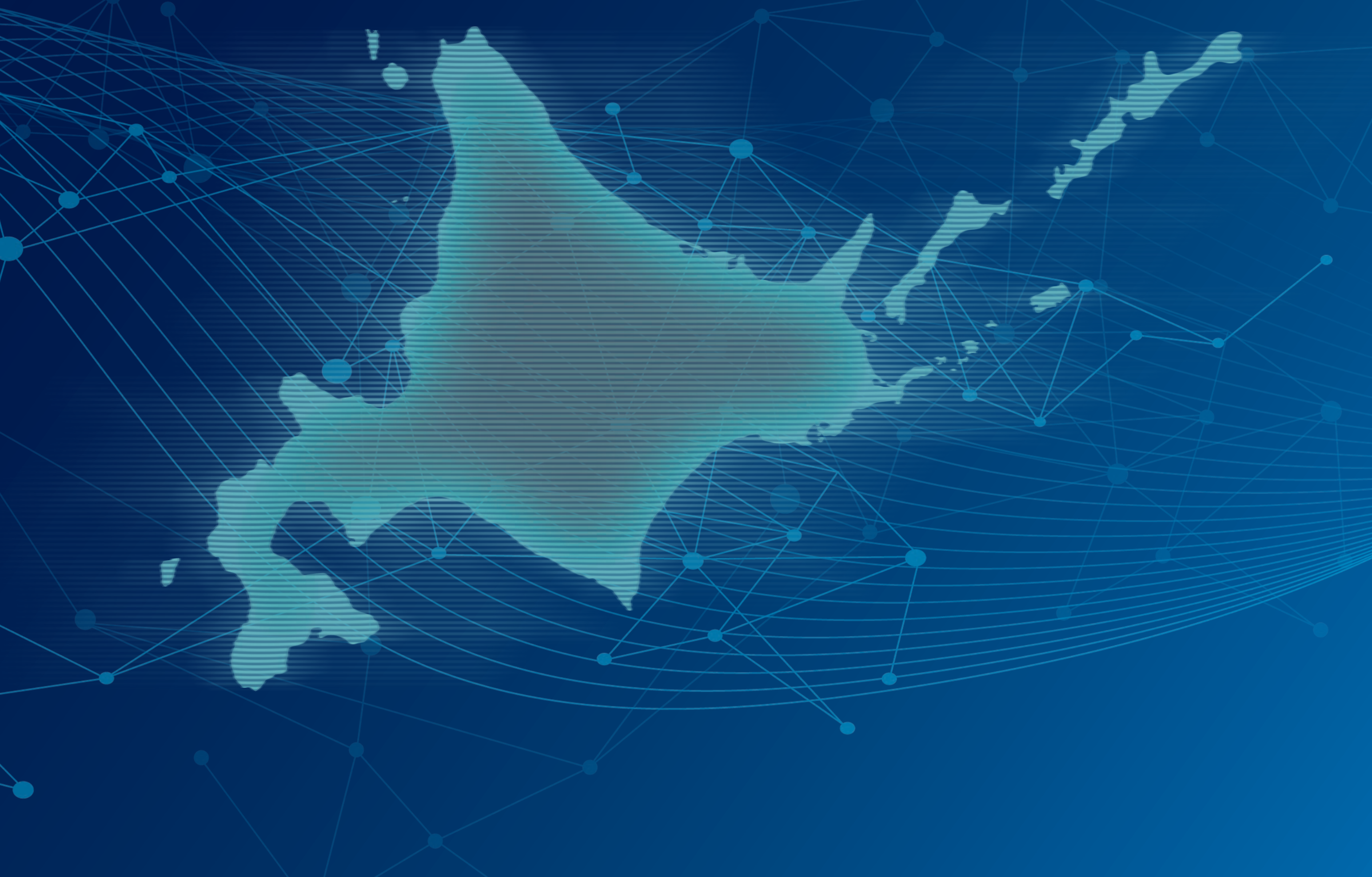

# INDEX

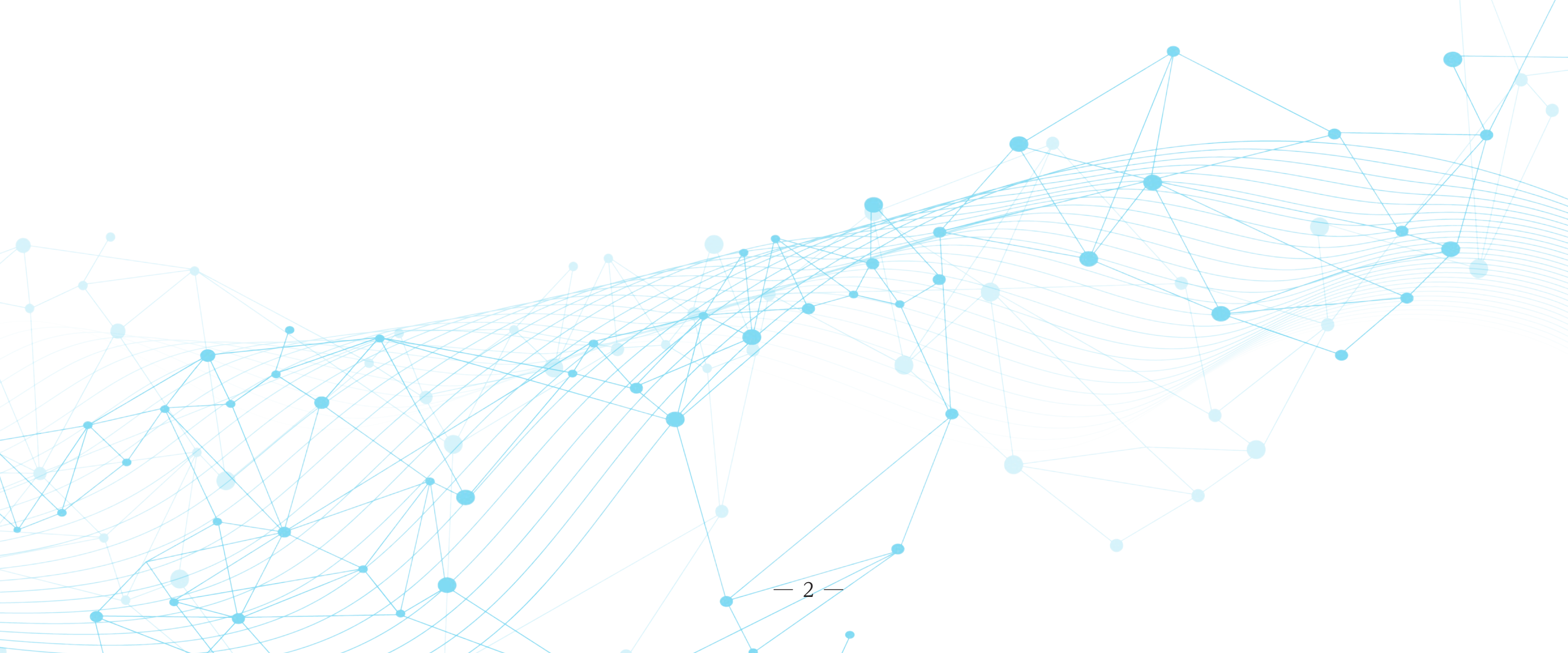

# 1. Purpose of This Report

This report summarizes the research outcomes achieved by the research group at Hokkaido University under the funding program
"Innovative Information and Communications Technology Research and Development (Commissioned Research)" of the National Institute of Information and Communications Technology (NICT).

Although this report is primarily written for specialists in the fields of optics and computational science, the content has been concisely organized to ensure accessibility and interest for a broader readership. For more detailed technical discussions, readers are referred to the relevant publications listed in the research achievements.

It is our sincere hope that this report will serve as a useful reference for future research and development in holography.

Project Overview

Program Title:

Innovative Information and Communications Technology Research and Development (Commissioned Research),
National Institute of Information and Communications Technology (NICT)

Project ID:

06801 (Function-Realization-Oriented Program, General Category)

Research and Development Theme:

Research and Development of Highly Efficient Compression and Transmission Technologies for Holographic Communication in Beyond 5G Networks

Researchers at Hokkaido University:

Yuji Sakamoto, Takahiro Yamanoi, Seok Kang
(see the project history section for details)

Research Objective:

This project aims to establish highly efficient compression and transmission technologies for multimodal information—including holography—covering three of the five human senses: vision, hearing, and touch. The goal is to contribute to the effective use of radio spectrum and to the creation of immersive viewing experiences that characterize Beyond 5G systems. Specifically, the project targets compression to a bit rate of 1.5 Gbps or lower under Beyond 5G network assumptions, corresponding to an eightfold improvement in efficiency compared with conventional methods, and plans to demonstrate the world's first end-to-end real-time transmission of such data.

Project Period:

January 2023 to March 2026

Participating Institutions:

- KDDI Research, Inc.
- Hokkaido University
- Tokai National Higher Education and Research System, Nagoya University
- Kansai University
- Suwa University of Science
- Crescent Inc.

## 2. Holometric Video Streaming

Holography is regarded as an ideal three-dimensional display technology, as it is theoretically capable of reconstructing images so realistically that they are indistinguishable from the original captured objects. For this reason, television systems based on holographic technology (holographic TV, or Holo-TV) have long been expected to become practical. Since early investigations such as those conducted at IBM Bell Laboratories in 1966, numerous component technologies and system architectures have been proposed toward its realization. However, due to technical challenges arising from the enormous volume of hologram data and the limited resolution of display devices, practical implementation was considered to be a goal for the distant future. In recent years, however, advances in computing technology, communication technology, and electronic display devices have significantly improved the surrounding technological environment. In particular, the emergence of large-capacity communication systems envisioned for Beyond 5G has made the transmission of hologram data increasingly feasible.

As illustrated in Fig. 1, a Holo-TV system is expected to support a wide variety of holographic display devices used by individual users. Examples include holographic head-mounted displays (Holo-HMDs), tablet-type devices, bench-top displays, monitor-type displays, projector-based systems, and room-scale installations (Holo-rooms). Because these devices differ in terms of optical configurations and display specifications, different hologram data are required depending on the device architecture. Furthermore, in order to provide a highly immersive user experience, it is essential to present images that respond to the user's free movement, i.e., free-viewpoint rendering. In particular, support for six degrees of freedom (6DoF) in user motion is highly desirable.

However, since both the holographic display devices and the corresponding viewpoint positions vary widely among users, generating and transmitting hologram data tailored to each individual user at the broadcasting side would result in an enormous computational and communication load.

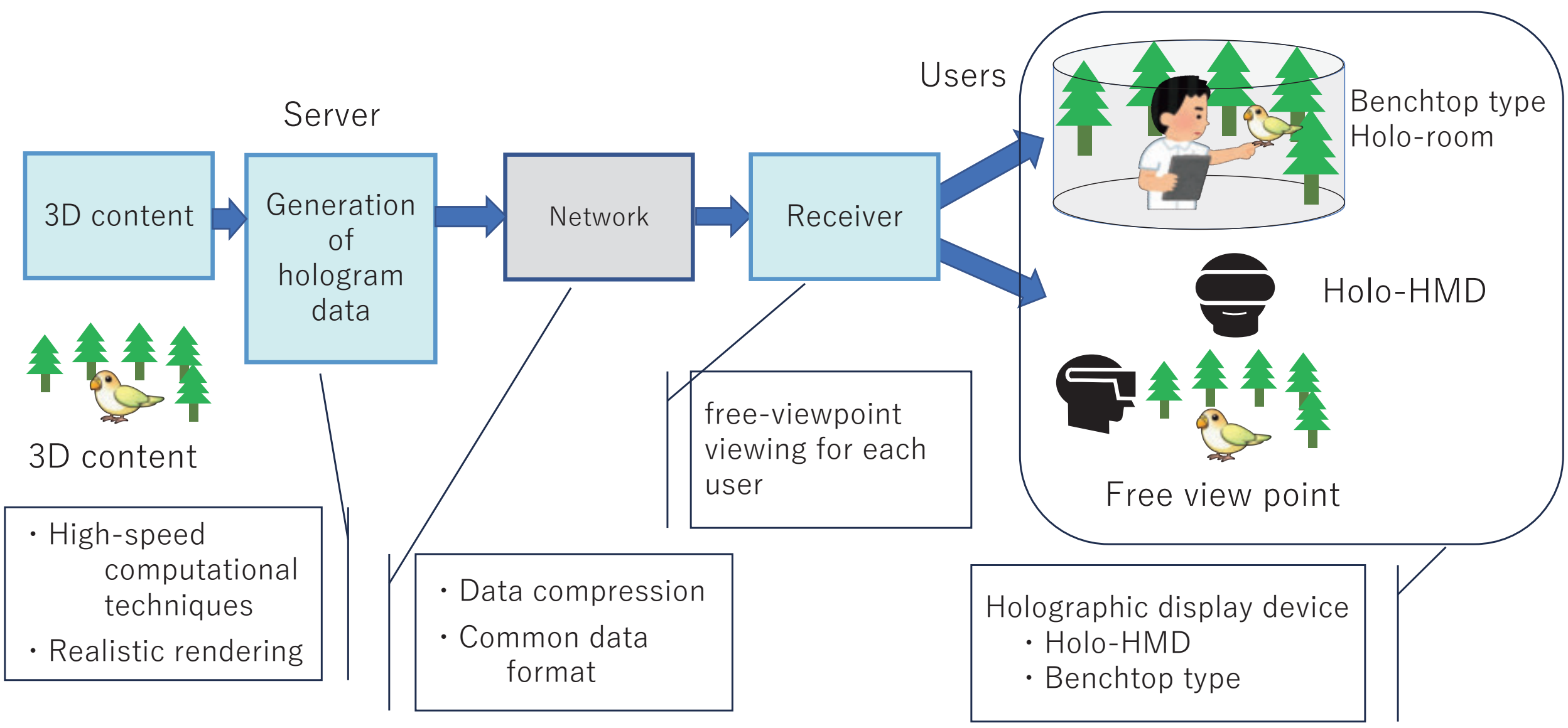

Fig. 1 Overall architecture of the Holometric Video Streaming (HVS) system, from 3D content capture at the broadcasting side to holographic display at the user side.

To address the issues of computational load and communication bandwidth in Holo-TV systems, we propose a system architecture referred to as Holometric Video Streaming (HVS) [40, 41]. As illustrated in Fig. 1, three-dimensional content is captured in a studio at the broadcasting side by recording the target objects, from which data suitable for holographic representation are generated. These data are then transmitted (or stored), and the content is viewed at the receiving side using holographic display devices.

In this system, instead of transmitting hologram data tailored to individual users, common data that are independent of the user's display device and viewpoint position are transmitted. At the user side, the received common data are converted into hologram data appropriate for each specific holographic display device.

To realize such a system, however, there remain several technical challenges that must be addressed.

- Generation of common data at the broadcasting side from captured 3D data
    - Methods for generating realistic holographic images
    - High-speed computational techniques
- Compression of hologram data with extremely large data volumes
- A common data format that realizes the following properties:
    - Independent free-viewpoint viewing for each user
    - Generation of hologram data tailored to each type of holographic display device
- Development of holographic display devices

This report presents the results of research on the elemental technologies required to realize the above objectives. Owing to the complexity of the proposed solutions and their application scenarios, an overview of the system is illustrated in Fig. 2.

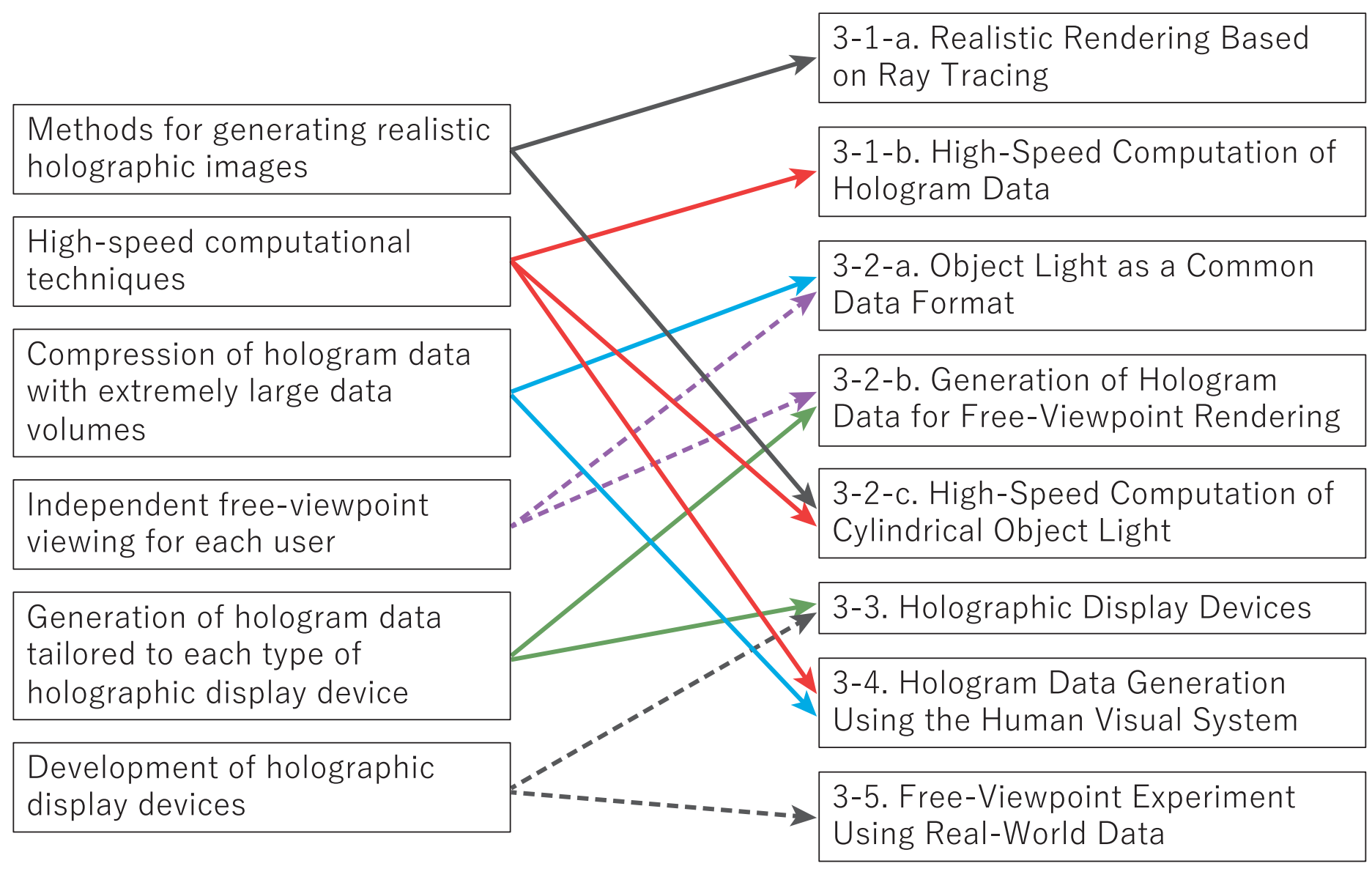

Fig. 2 Conceptual overview of the proposed HVS framework and the elemental technologies addressed in this study.

# 3. Research Results

## 3-1. Generation of Holographic Data at the Broadcasting Station

### 3-1-a. Realistic Rendering Based on Ray Tracing

The process of generating hologram data from 3D data obtained as polygonal or point-cloud representations is referred to as computer-generated holography (CGH). In this process, realistic rendering techniques that account for hidden surface removal and surface material properties are critically important for obtaining reconstructed images with a strong sense of depth. However, in CGH, computational algorithms that sufficiently incorporate such rendering effects have not yet been fully established. Therefore, this study investigates a realistic rendering method for CGH based on ray tracing.

In CGH computation using ray tracing, as illustrated in Fig. 3-1, light rays are traced from each point constituting the object toward the hologram plane. If a ray intersects another object along its path, the light from that point is assumed not to reach the hologram plane and is thus treated as belonging to a hidden surface. In other words, the corresponding point is regarded as being occluded and is not visible at the hologram plane. Furthermore, in the presence of mirrors, the ray paths must be determined while accounting for reflections at mirror surfaces.

Previous studies have realized hidden surface removal, surface reflection modeling, and reflections by planar mirrors. In contrast, real-world environments contain many curved reflective surfaces, such as metallic doorknobs, automobile exteriors, glass containers, and curved traffic mirrors. In this project, we investigated an algorithm for generating surrounding scenes reflected by curved mirrors.

For representing curved surfaces, Bézier surfaces, which are widely used in the field of computer graphics, were adopted. In the proposed algorithm, reflection points and optical paths on Bézier surfaces are determined using Bézier clipping, and hologram data are calculated based on these results [16]. This approach enables the generation of reconstructed images that include surrounding scenes reflected by curved surfaces. Moreover, the method supports motion parallax corresponding to viewpoint movement, and it was confirmed that natural three-dimensional images can be observed.

However, this method suffers from a significant computational load, requiring several hours to generate hologram data with a resolution of 2K (1920 × 1080 pixels). To address this issue, we proposed a fast computation method in which Bézier surfaces are subdivided into small planar segments, achieving a speedup of more than 30 times (Fig. 3-2) [13, 22, 25].

Furthermore, by introducing a method based on Fermat's principle for optical path determination, an additional acceleration of approximately 100 times was achieved, enabling hologram data generation on the order of seconds [31, 42]. As a result, reflection rendering involving multiple reflective surfaces, which had previously been difficult due to computational constraints, was successfully realized. Fig. 3-3 shows a reconstructed image in which a rabbit is placed in front of the observer, with the first mirror located behind the observer and the second mirror positioned behind the rabbit. In the photograph, the rabbit and two images produced by double reflections can be observed. The difference in depth between these images indicates that accurate reflection rendering has been achieved.

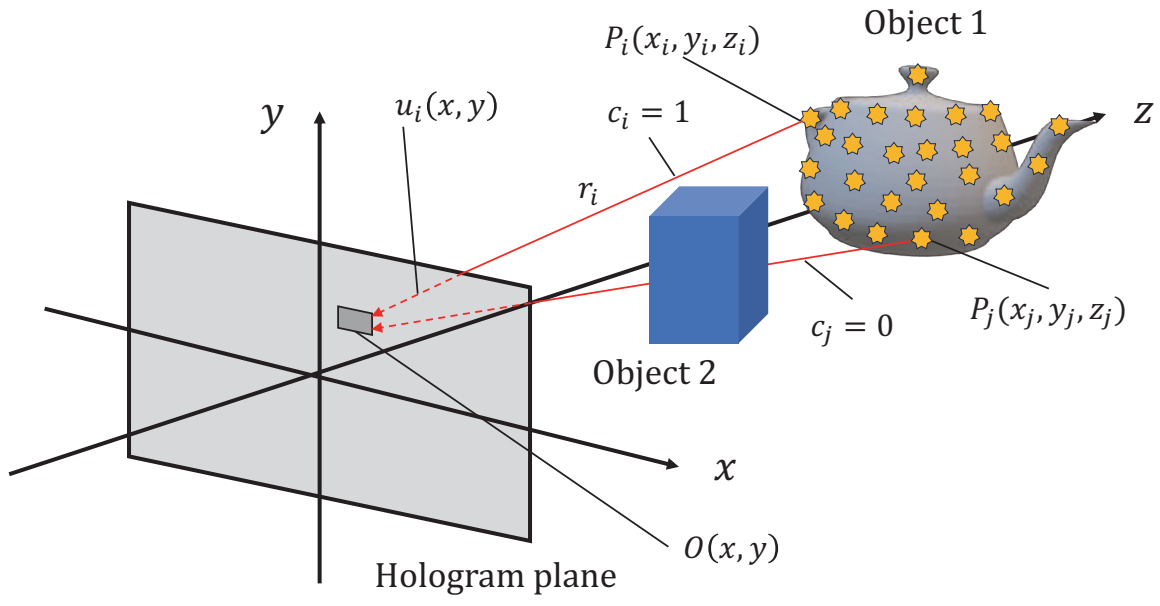

Fig. 3-1. Hidden surface removal using the point light source method and the ray tracing method.

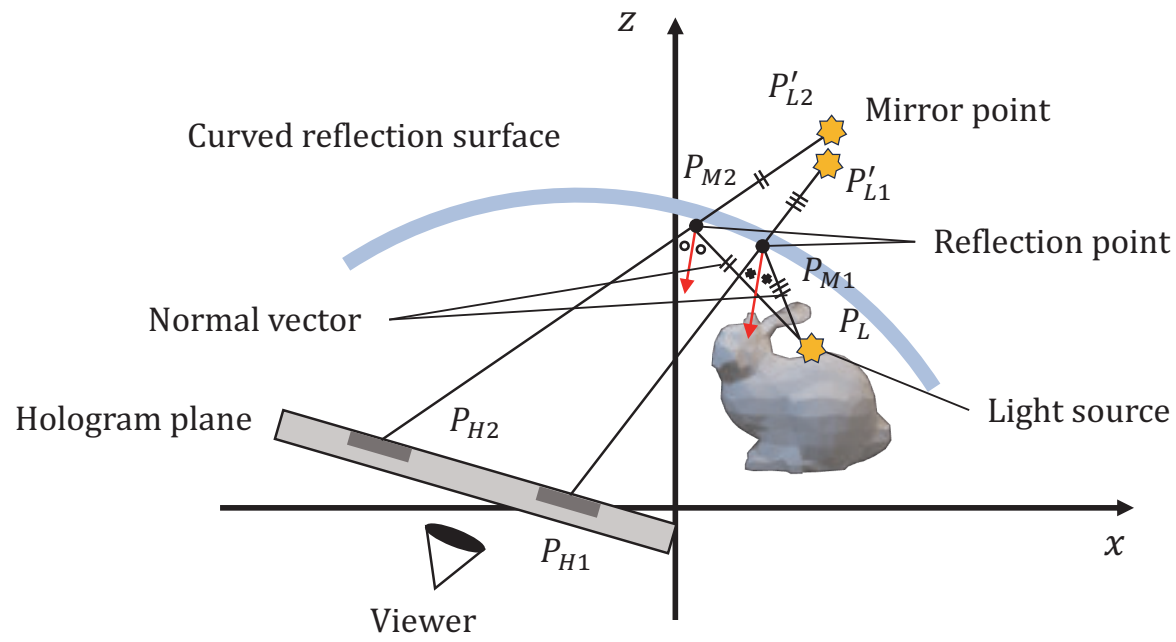

Fig. 3-2. Principle of reflection calculation on curved surfaces.

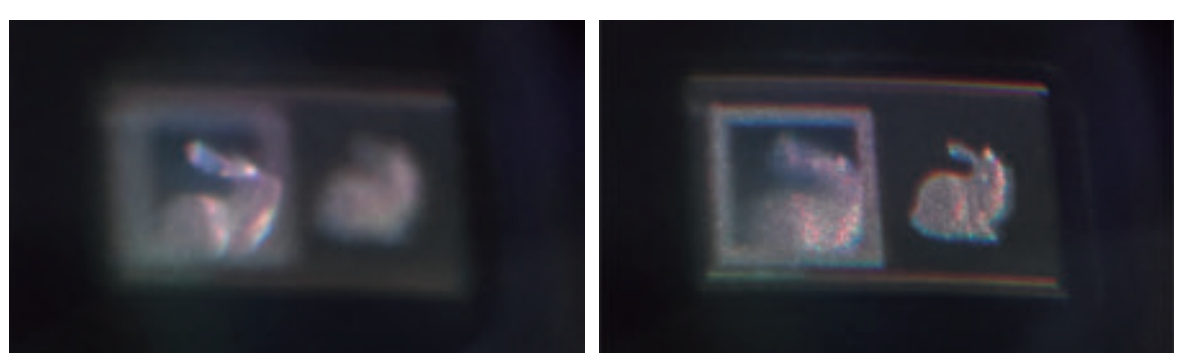

Fig. 3-3. Reconstructed images reflected by two mirrors. The left image focuses on the real image, while the right image focuses on the image produced by double reflections.

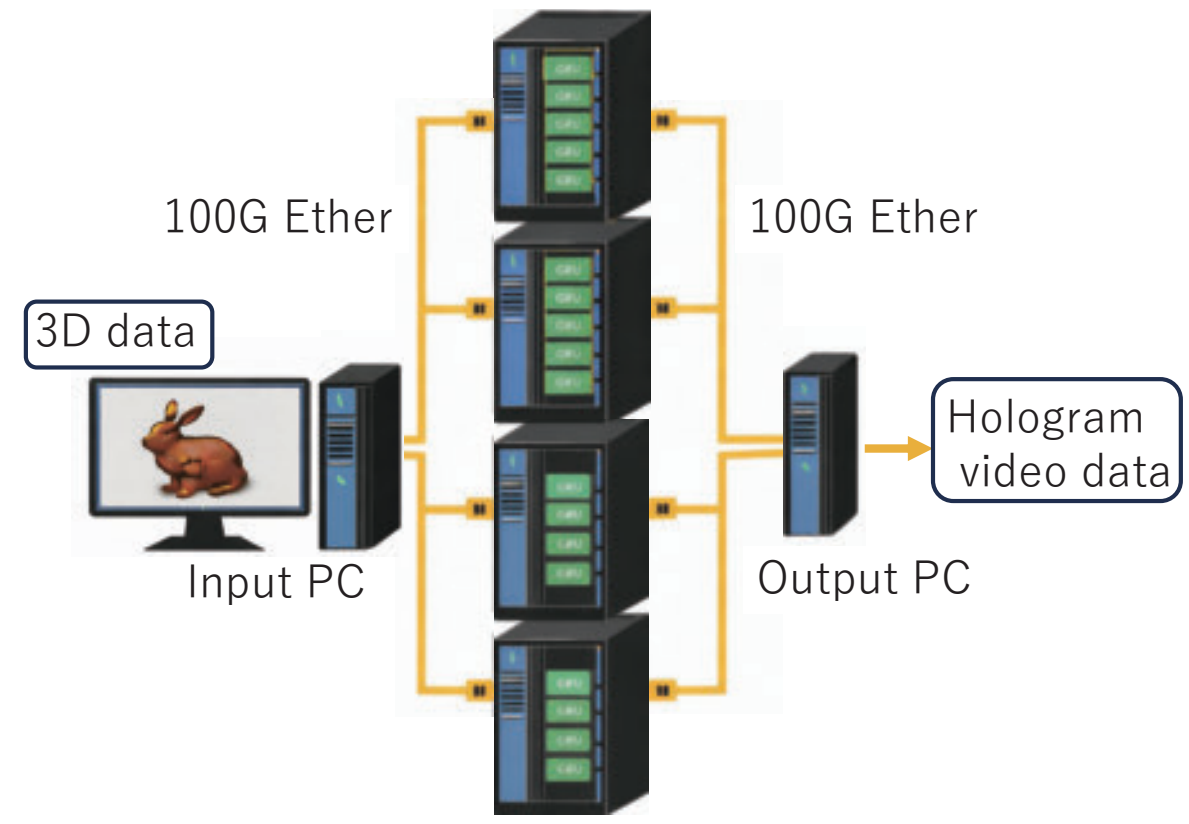

Fig. 3-4. Configuration of the GPU server system.

### 3-1-b. High-Speed Computation of Hologram Data — Parallel Computation Using a GPU Cluster —

The generation of hologram data requires an enormous amount of computation, and high-speed computation techniques are indispensable for achieving real-time display. To address this challenge, many research groups have investigated acceleration approaches from both algorithmic and hardware perspectives. In these approaches, utilizing the computational power of high-performance computing systems is an effective solution. Large-scale computing systems can be deployed at the broadcasting station.

Hologram data computation is well suited to parallel processing, and efficient acceleration can be achieved by employing a large number of computational cores. Accordingly, in this project, a cluster computing system utilizing multiple Graphics Processing Units (GPUs) was constructed [6, 8, 30, 36].

The constructed computing system consists of four servers equipped with a total of 18 GPUs (NVIDIA RTX 4090), as shown in Fig. 3-4. The servers are interconnected via a 100-Gbps Ethernet network. Since each GPU contains several thousand computational cores, the entire system comprises approximately 90,000 cores and provides a theoretical peak performance of 2 PFLOPS.

To achieve highly efficient processing in this computational environment, a high-speed computation algorithm was developed by combining the following three techniques [10, 11]:

- The CGH computation was divided into three stages—preprocessing (point light source placement), main processing (object light computation), and postprocessing (reference light and interference fringe generation)—and these stages were pipeline-processed.
- The hologram plane was spatially divided, and a parallel computation algorithm utilizing GPU cores was introduced.
- The 18 GPUs processed different frames in parallel.

The computation algorithm employed the ray tracing method described in the previous section. The hologram size was set to 2K, and real-time video generation was achieved under conditions with a relatively small number of point light sources. In the video generation experiments, complex motions were tested, including physical simulations such as object falling and bouncing, as well as a puzzle game in which a ball is placed into a hole. It was confirmed that these motions were successfully displayed in the optically reconstructed images.

Fig. 3-5 shows an example of a reconstructed video of a physical simulation in which an iron ball is dropped onto one side of a seesaw with a rabbit on the opposite side. The rabbit is observed to jump upward due to the impact of the falling iron ball. In this experiment, the rabbit-and-seesaw scene was computed using 10,000 point light sources over 120 frames. The relationship between the number of GPUs and the computation time is shown in Fig. 3-6. When all 18 GPUs were utilized, the total generation time for all frames was 4.661 s, corresponding to 25.8 frames per second (FPS), thereby achieving

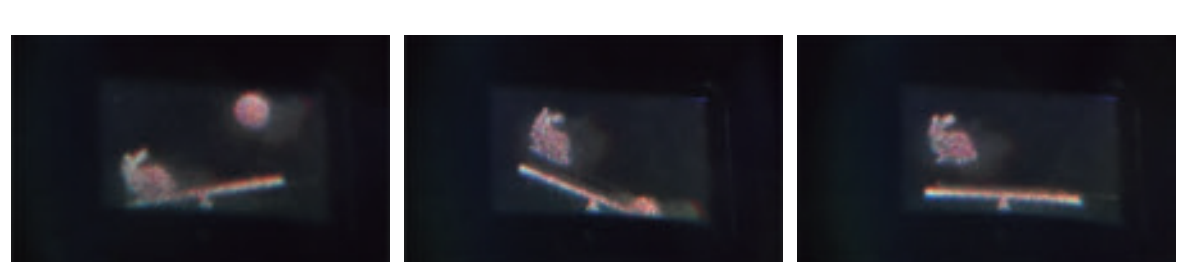

Fig. 3-5. Reconstructed video sequence (rabbit and seesaw). Time progresses from left to right.

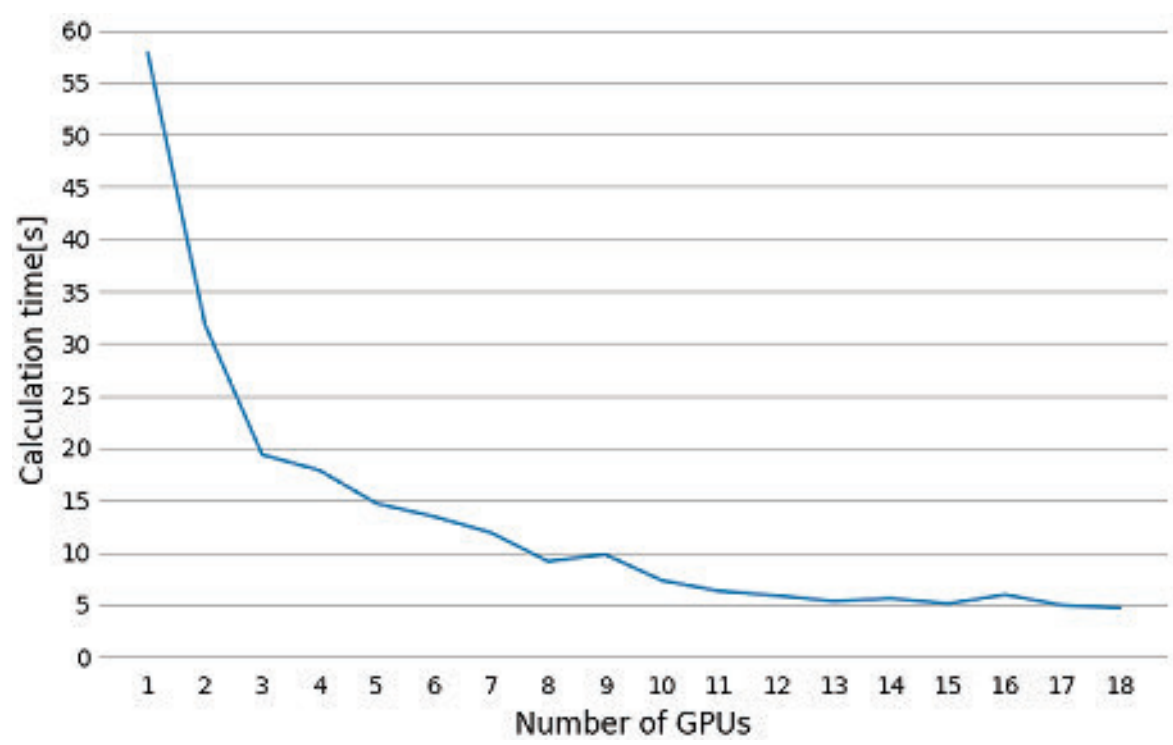

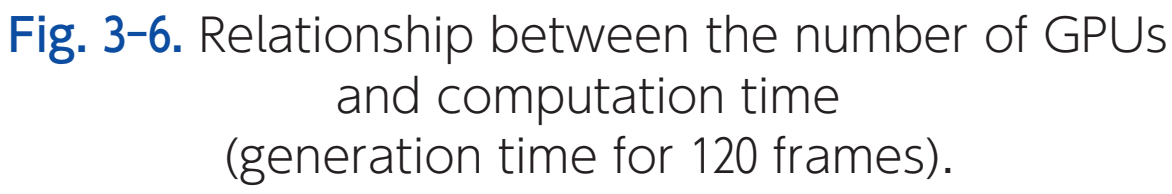

Fig. 3-6. Relationship between the number of GPUs and computation time (generation time for 120 frames).

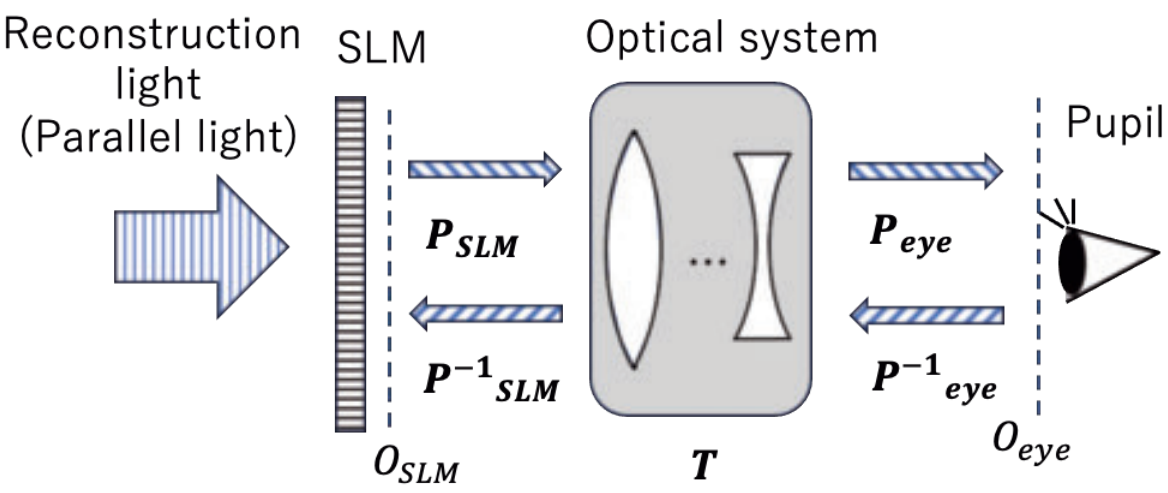

Fig. 3-7. Optical concept of a holographic display device.

quasi-real-time performance.

Through these techniques, large-scale hologram data can be generated at high speed, providing the computational performance required for the generation of dynamic holographic imagery.

## 3-2. Common Data Format and Generation of Hologram Data

### 3-2-a. Object Light as a Common Data Format

In the HVS, the hardware configuration, device specifications, and optical system design of holographic display devices differ among users. Accordingly, the hologram data also vary depending on the device, and the computational algorithms used to generate the hologram data are not identical. To address this issue, this study proposes a method in which data are transmitted using a common data format and then converted on the receiver side into hologram data suitable for each device [28, 34, 40].

In general, a holographic display device consists of a spatial light modulator (SLM) and an optical system, as shown in Fig. 3-7. To obtain a reconstructed image, hologram data are displayed on the SLM, which is illuminated by collimated light as the reconstruction light. The light $O_{SLM}$ generated by the SLM propagates toward the optical system (propagation operator $P_{SLM}$) and undergoes an optical transformation $T$. The light then propagates to the pupil (propagation operator $P_{eye}$), and the reconstructed image is perceived by observing the optical wave $O_{eye}$ at the pupil position. This relationship can be expressed as

$$O_{eye}=P_{eye}\ T\ P_{SLM}\ O_{SLM}, \qquad (1)$$

Since an optical system generally consists of multiple lenses and spherical mirrors, the transformation $T$ represents a combined transformation that includes the effects of each optical element as well as the propagation between them.

To ensure that the same reconstructed image can be observed by users regardless of the type of holographic display device, it is sufficient for the optical wave $O_{eye}$ at the pupil position to be identical. To achieve this condition, the light $O_{SLM}$ after the SLM must be determined while taking into account the optical system of each display device. From Eq. (1), $O_{SLM}$ can be expressed as

$$O_{SLM}=P^{-1}{}_{SLM}\ T^{-1}\ P^{-1}{}_{eye}\ O_{eye} \qquad (2)$$

Here, $P^{-1}$ and $T^{-1}$ denote the inverse transformations of $P$ and $T$, respectively, corresponding to back propagation. Equation (2) indicates that, starting from the same $O_{eye}$, a unique $O_{SLM}$ corresponding to the optical system of each user's holographic display device can be obtained. Therefore, $O_{eye}$ can be treated as device-independent common data and is suitable as a transmission format in broadcast-type holographic systems.

Fig.3-8 shows reconstructed images obtained using a holographic display device. Fig. 3-8(a) shows a reconstructed image generated from the object light $O_{eye}$ at the viewpoint, representing the ideal image that should be observed; a star-shaped image is clearly visible. Next, a holographic display device incorporating a single convex lens is considered. Fig. 3-8(b) shows the result obtained when hologram data generated by converting $O_{eye}$ using the proposed method are observed with a holographic display device without the lens, where significant blurring is observed. In contrast, Fig. 3-8(c) shows the result

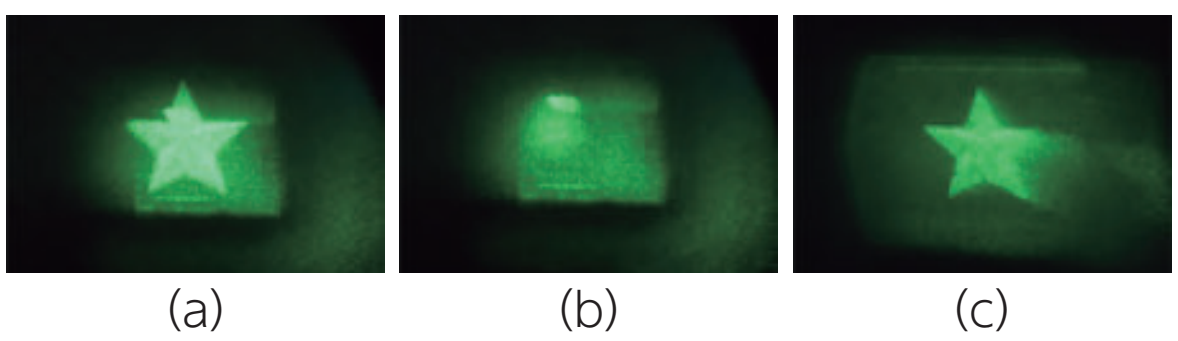

**Fig. 3-8.** Reconstructed images obtained using the proposed method.

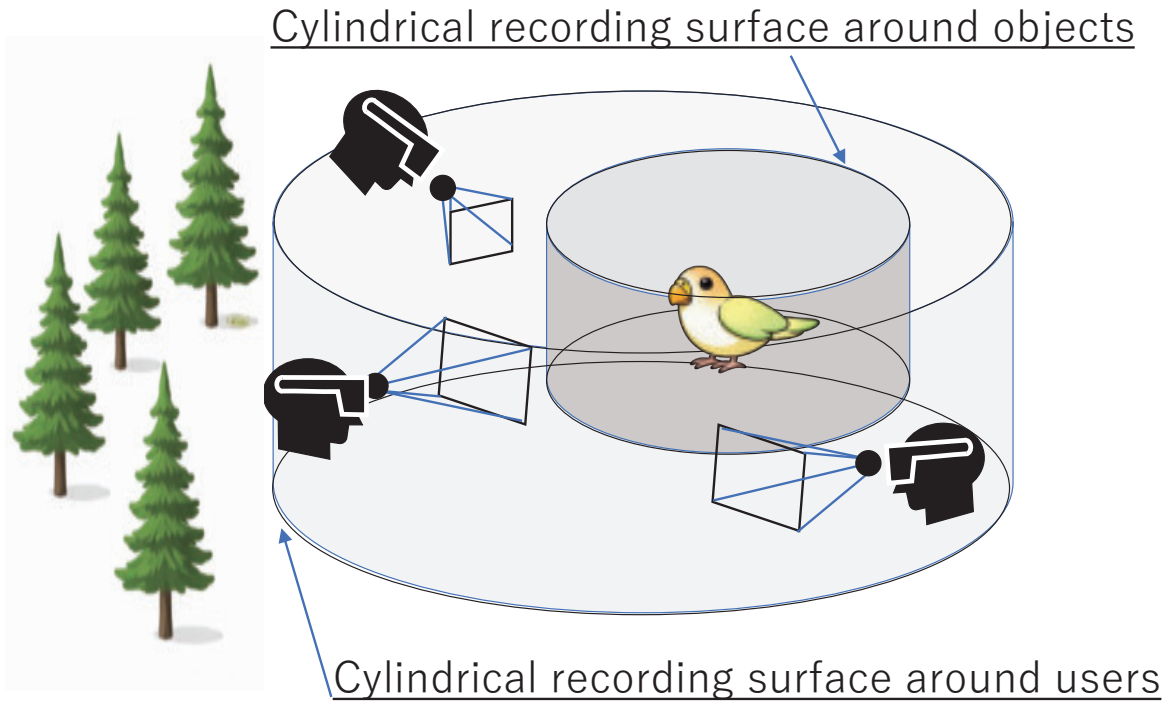

**Fig. 3-9.** Propagation from the cylindrical object-light recording surface to the hologram plane.

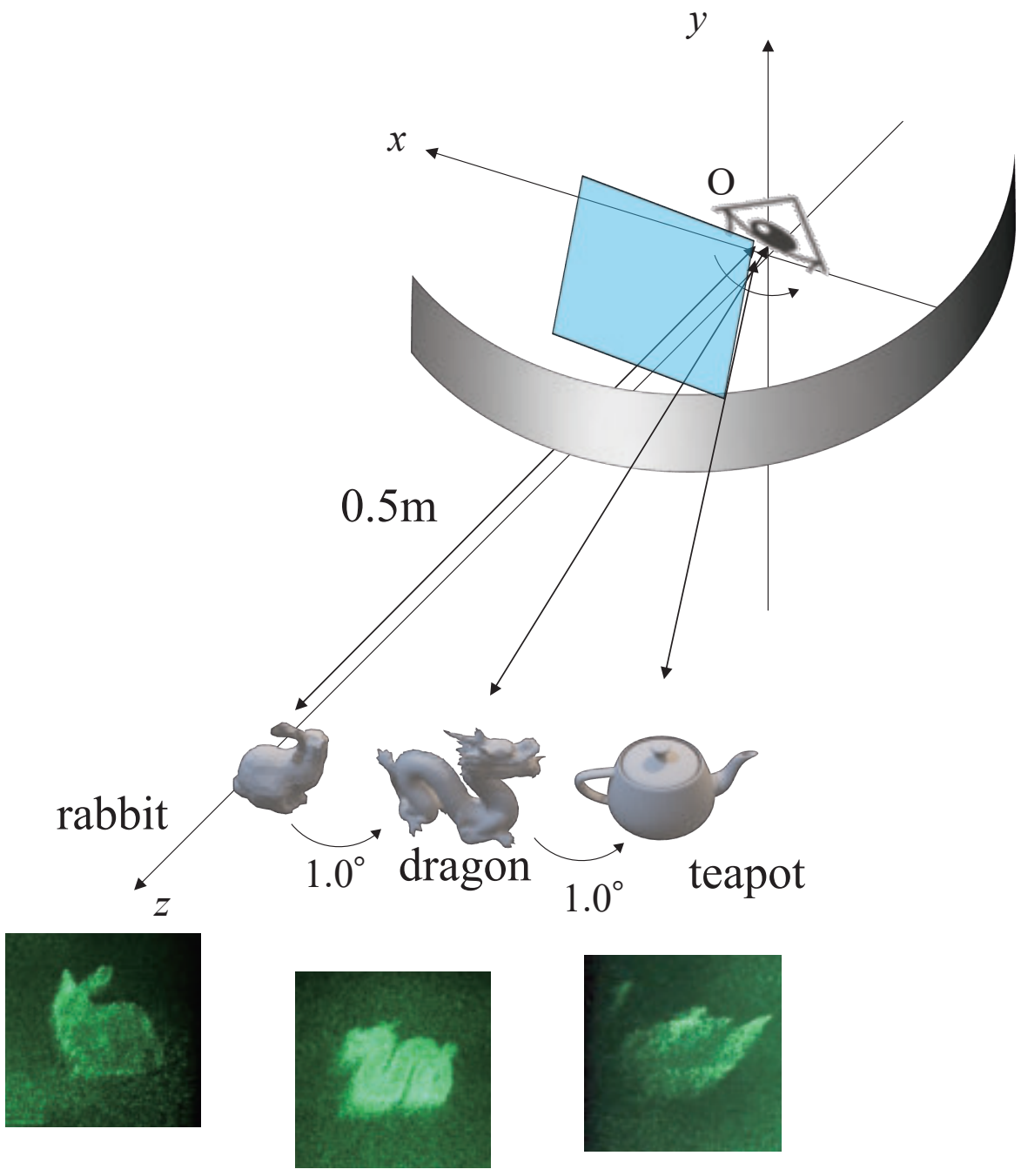

**Fig. 3-10.** Holographic panoramic images obtained by rotating the user's head.

obtained using a holographic display device with the lens, where the star-shaped image is observed clearly, similar to Fig. 3-8(a).

As described above, by employing the proposed common data format, it becomes possible to support a wide variety of holographic display devices while significantly reducing both computational and communication loads.

### 3-2-b. Generation of Hologram Data for Free-Viewpoint

Allowing each user to observe reconstructed images from an independent free viewpoint is essential for providing a highly immersive user experience. One possible approach to achieving this is to feedback each user's viewpoint information to the broadcasting station and generate hologram data tailored to each individual user at the broadcaster side. However, this approach imposes a substantial computational load on the broadcasting station and also requires the transmission of user-specific data over the communication network. As a result, this method suffers from scalability issues.

Accordingly, this study investigates a method that employs a cylindrical recording surface capable of recording wavefront data [15]. As illustrated in Fig. 3-9, light emitted from an object is recorded over the entire circumference of the cylindrical surface. By transmitting this information from the broadcasting station to a large number of users, the wavefront information at each user's hologram plane can be obtained by computing the propagation from the cylindrical surface to the user's hologram plane. In this manner, images corresponding to each user's viewpoint can be independently generated, enabling free-viewpoint holography with six degrees of freedom (6DoF).

By placing a cylindrical recording surface around the object, the object can be observed from various directions. Furthermore, by constructing a cylindrical recording surface around the user, panoramic images can also be displayed. Fig. 3-10 shows reconstructed images obtained by rotating a holographic HMD while the viewpoint is located at the center of the cylindrical surface that records the object light. It can be confirmed that different objects appear in the field of view depending on the observation direction, demonstrating panoramic image presentation.

### 3-2-c. High-Speed Computation of Cylindrical Object Light

When object light on a cylindrical recording surface is directly computed using the point light source method, the computational cost becomes extremely large. This problem becomes even more severe when realistic rendering is considered, as additional computations such as hidden surface removal and reflection calculations are required.

This study developed an algorithm for efficiently computing object light on a cylindrical recording surface from object light that has already been realistically rendered and recorded on a planar surface.

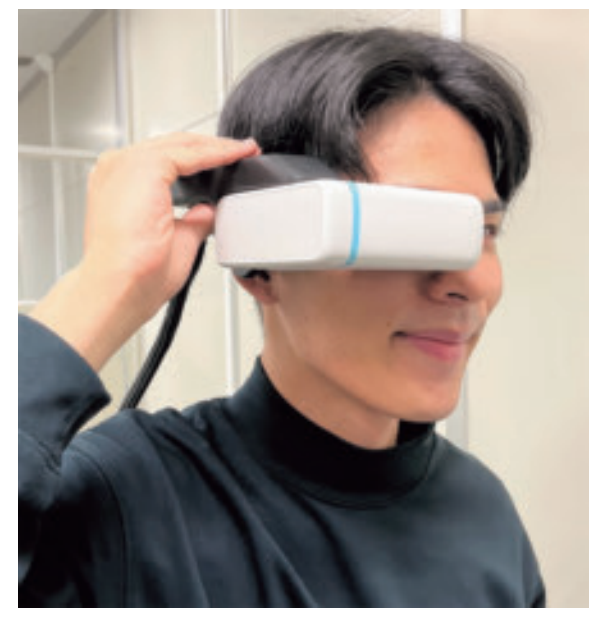

Fig. 3-11. Scene of a user operating the Holo-HMD.

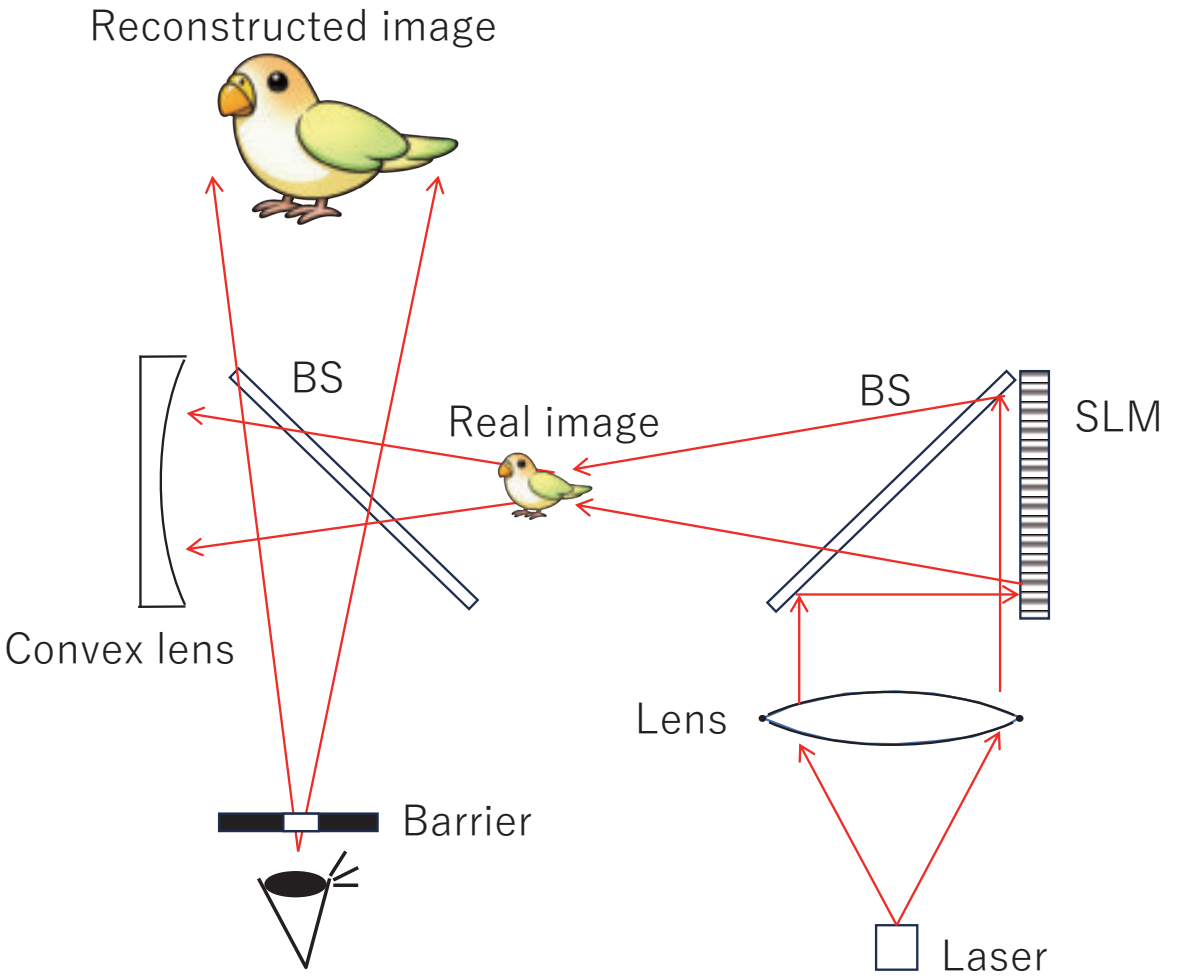

Fig. 3-12. Optical system of the Holo-HMD.

The proposed method consists of the following three steps [18]:

1. A planar recording surface is placed near the object, and object light that incorporates realistic rendering effects from the object is recorded.
2. Another planar recording surface is placed at the center of the cylinder, and object light is recorded by performing back propagation of the wavefront from the previously recorded surface.
3. Wavefront propagation from the object light located at the center is computed toward the cylindrical recording surface.

Using this approach, object light corresponding to realistically rendered images can be computed more than 10,000 times faster than with conventional methods.

## 3-3. Holographic Display Devices

### 3-3-a. Holographic Head-Mounted Display (Holo-HMD)

The developed Holo-HMD employs an optical system based on previous studies and enables full-color reconstruction using a single RGB laser light source (Figs. 3-11 and 3-12). In addition, a convex mirror is used to expand the field of view (FOV), achieving a relatively wide display area despite its head-mounted configuration. The theoretical FOV is 18.2°, while the experimentally measured FOV is 16.4°. The mounted spatial light modulator (SLM) has a resolution of 2K with a pixel pitch of 4.5 μm. The external dimensions of the device are 125 × 130 × 40 mm, resulting in a compact form factor with an optical design suitable for AR displays. A motion sensor for measuring head rotation is also integrated into the device [4, 9, 20, 24, 43].

In VR/AR applications, it is essential to generate holograms in real time in response to head movements. However, CGH requires an enormous amount of computation, and the limited computational resources available on the user side make real-time processing difficult. To address this issue, we propose a system architecture in which a large-scale computing system is deployed at the broadcasting station to generate common data, while hologram data are generated on the user side using lightweight, low-complexity computational algorithms.

The algorithm in the user side performs phase compensation based on optical path differences according to the user's posture angle (Fig. 3-13). In conventional point light source methods, the computational cost increases in proportion to the number of point light sources, making real-time hologram updates difficult. In contrast, once the object light is given, the proposed method exhibits a computational cost that is nearly independent of the number of point light sources, allowing the processing to be completed in approximately constant time. As a result, rapid updates of CGH become possible. Experimental results confirmed that the reconstructed image follows head rotations and supports six degrees of freedom (6DoF). Furthermore, the update rate exceeded 30 frames per second (FPS), achieving real-time operation. An example of a reconstructed image is shown in Fig. 3-14.

From these results, it was confirmed that the proposed Holo-HMD enables AR display that follows the user's posture angle and achieves real-time hologram updates. In addition, the developed system satisfies the major requirements for practical HMDs, including support for six degrees of freedom (6DoF), full-color display, and a head-mountable size and weight, thereby demonstrating its effectiveness toward the practical realization of holographic HMDs.

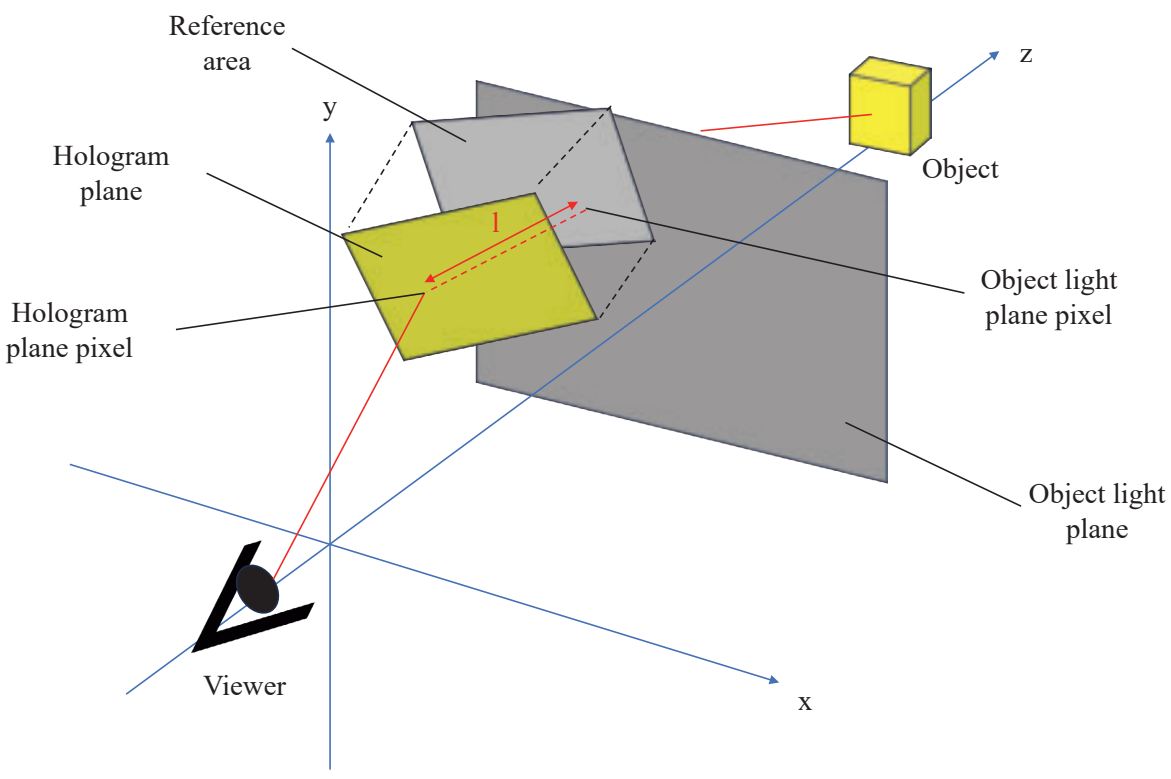

Fig. 3-13. Conceptual diagram of the phase compensation method.

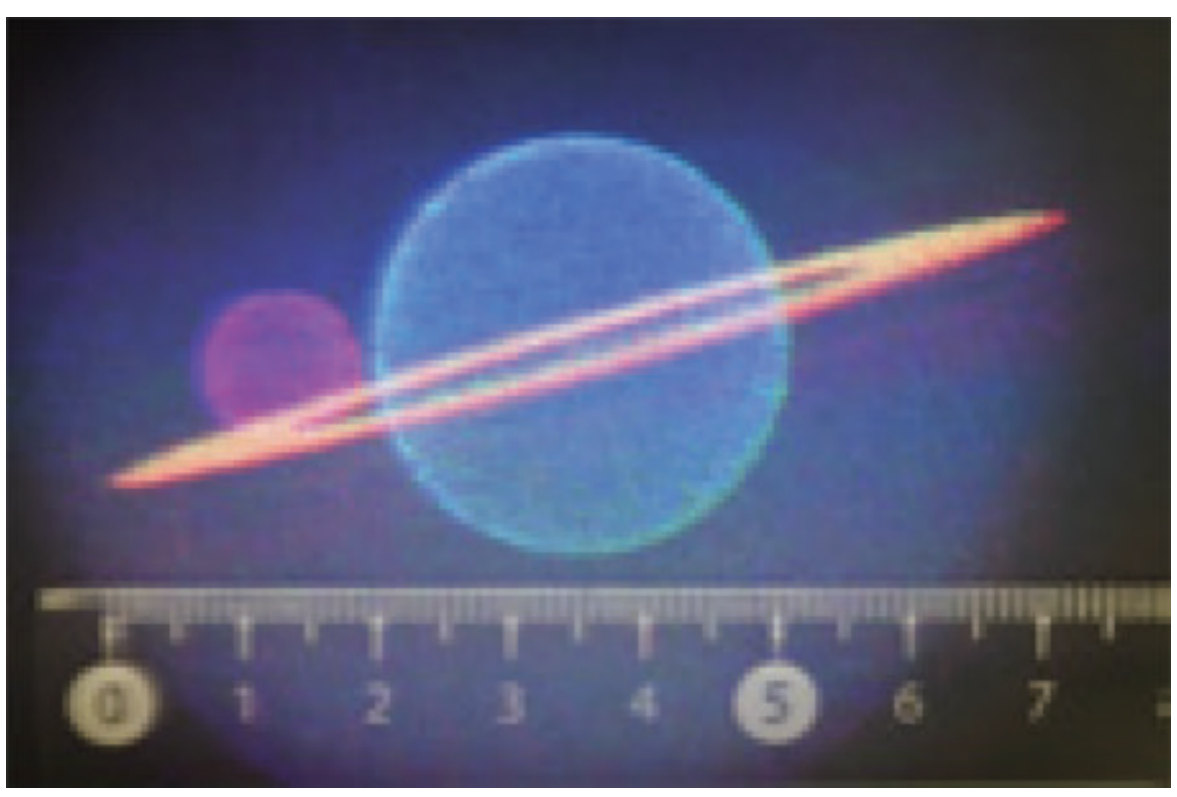

Fig. 3-14. Reconstructed image with the Holo-HMD.

### 3-3-b. Holographic Head-Mounted Projector (Holo-HMP)

HMDs have the drawback of obstructing the user's naked-eye field of view and limiting visibility of the real world. To address this issue, this study proposes a method for expanding the field of view using a head-mounted projector (HMP) [5, 19, 23].

An HMP is a display system in which an optical system is mounted on the user's head and used as a projector, enabling object images to be projected over a wide area. In this study, a configuration was adopted in which a concave mirror is placed in front of the user, and reconstructed images projected from the HMP are magnified by the concave mirror (Figs. 3-15 and 3-16). With this configuration, holographic images can be presented without blocking the user's field of view, unlike conventional HMDs, thereby allowing holographic visualization while preserving visibility of the real environment.

Furthermore, in this study, the holographic reconstructed image is dynamically updated according to the user's posture angle, particularly the head roll angle, in order to maintain visual consistency. An accelerometer is utilized to detect the user's posture, and the phase compensation algorithm also adopted in the Holo-HMD is applied for real-time processing. As a result, even when the user's head is tilted to the left or right, the holographic image is corrected so that it appears consistent with real-world objects, thereby achieving a more natural three-dimensional display.

From these results, it is demonstrated that the proposed HMP approach can expand the field of view of holographic display devices without obstructing the user's vision, while also correcting reconstructed images according to the user's posture angle.

### 3-3-c. Portable Dedicated Hologram Data Generation Device

Conventionally, the generation of hologram data requires a PC equipped with GPUs, resulting in systems that are large and heavy and thus unsuitable for portable applications. To address this issue, a portable hologram data generation device, as shown in Fig. 3-17, was developed [7, 12, 21, 39]. The device is a dedicated computation system based on a field-programmable gate array (FPGA) and is designed specifically to generate hologram data from the cylindrical object light described above.

An XCVU9P-L2FLGA2104E FPGA (logic cells: 2,586,150) was employed (Fig. 3-18). Since hologram computation can be performed independently for each pixel, two types of circuit architectures were implemented and evaluated: pixel-wise parallel processing and a pipelined structure. Experimental results confirmed that hologram data generation synchronized with head rotation was correctly performed in both configurations. In terms of computational speed, the pipelined architecture achieved higher performance, generating hologram data with a resolution of 128 × 128 pixels in 0.68 ms. This corresponds to a quasi-real-time update rate of 11 frames per second (FPS) when scaled to a 2K SLM used in the Holo-HMD. Although further acceleration is required, these results demonstrate the feasibility of a portable dedicated hologram data generation device.

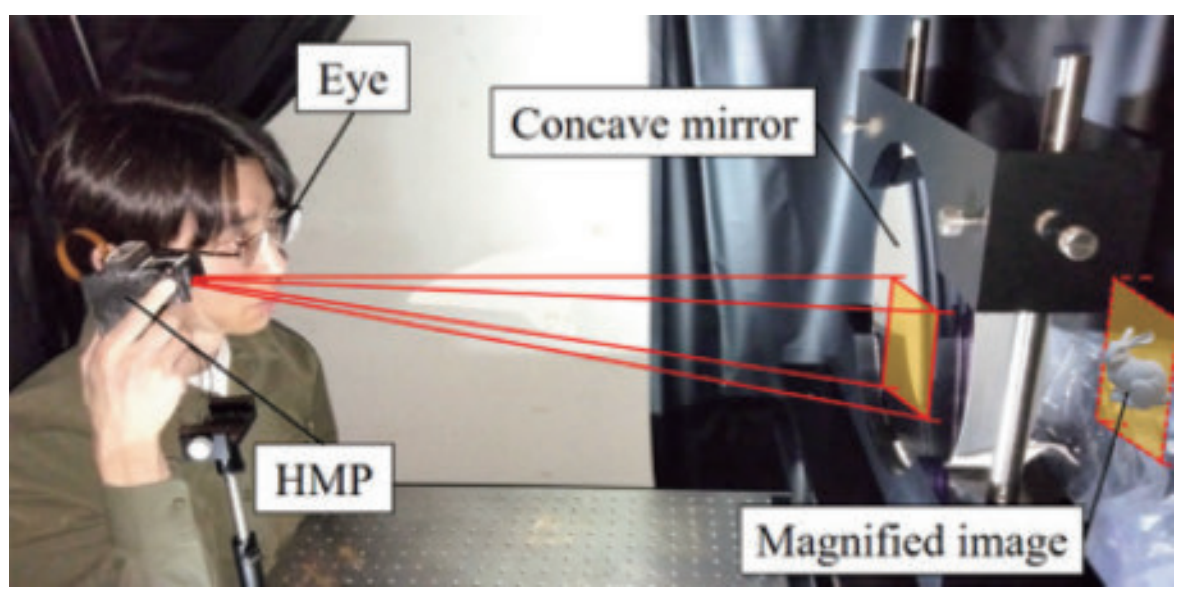

Fig. 3-15. Scene of a user operating the Holo-HMP.

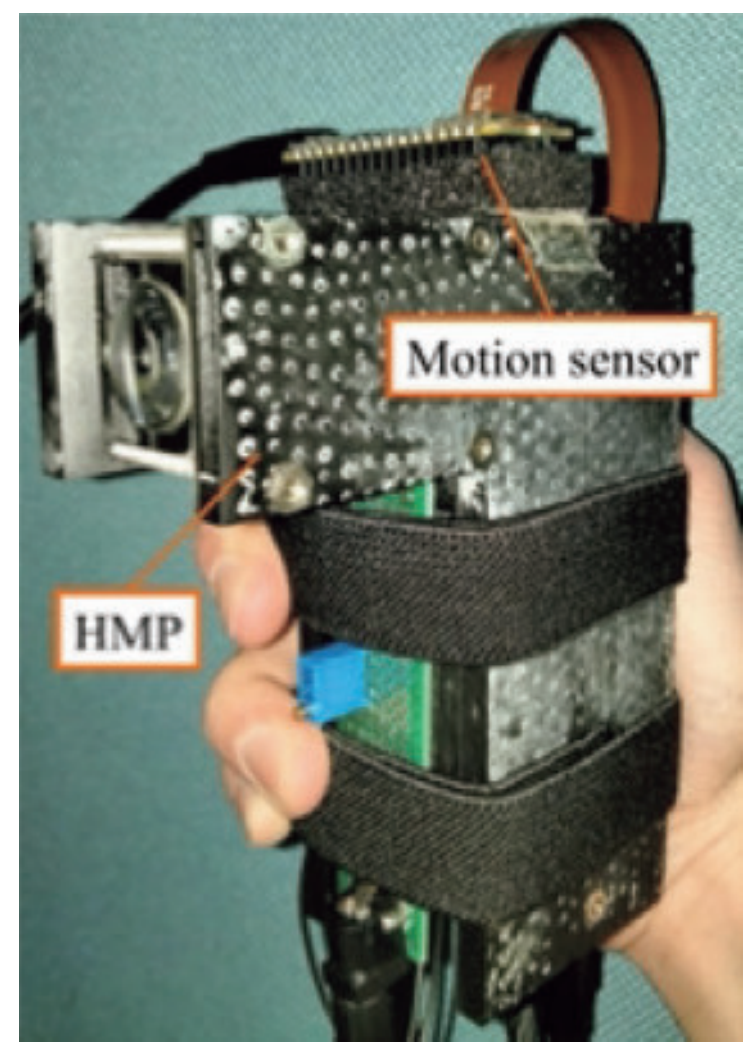

Fig. 3-16. Head-mounted unit of the Holo-HMP.

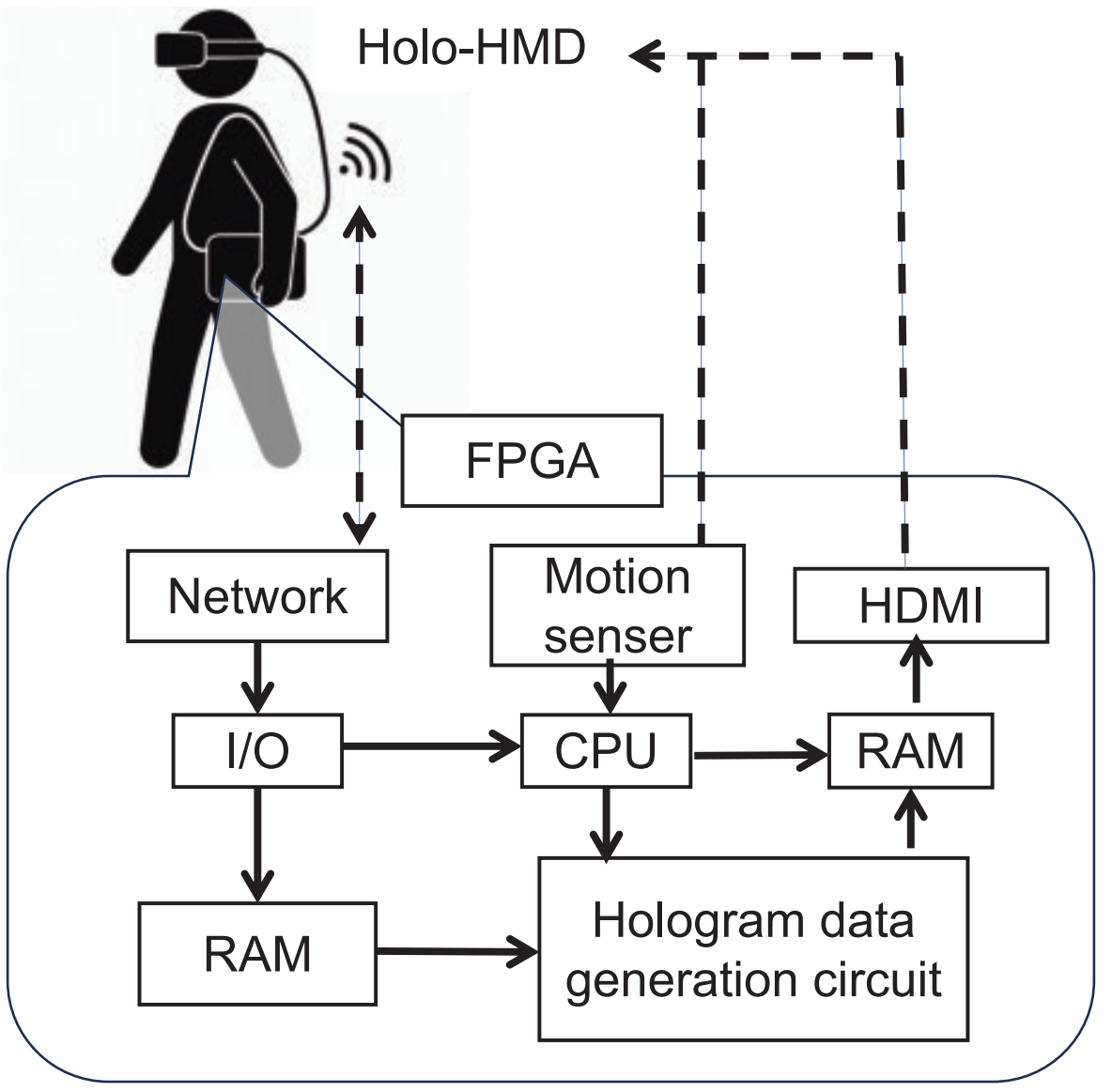

Fig. 3-17. Basic configuration of the portable dedicated hologram data generation device.

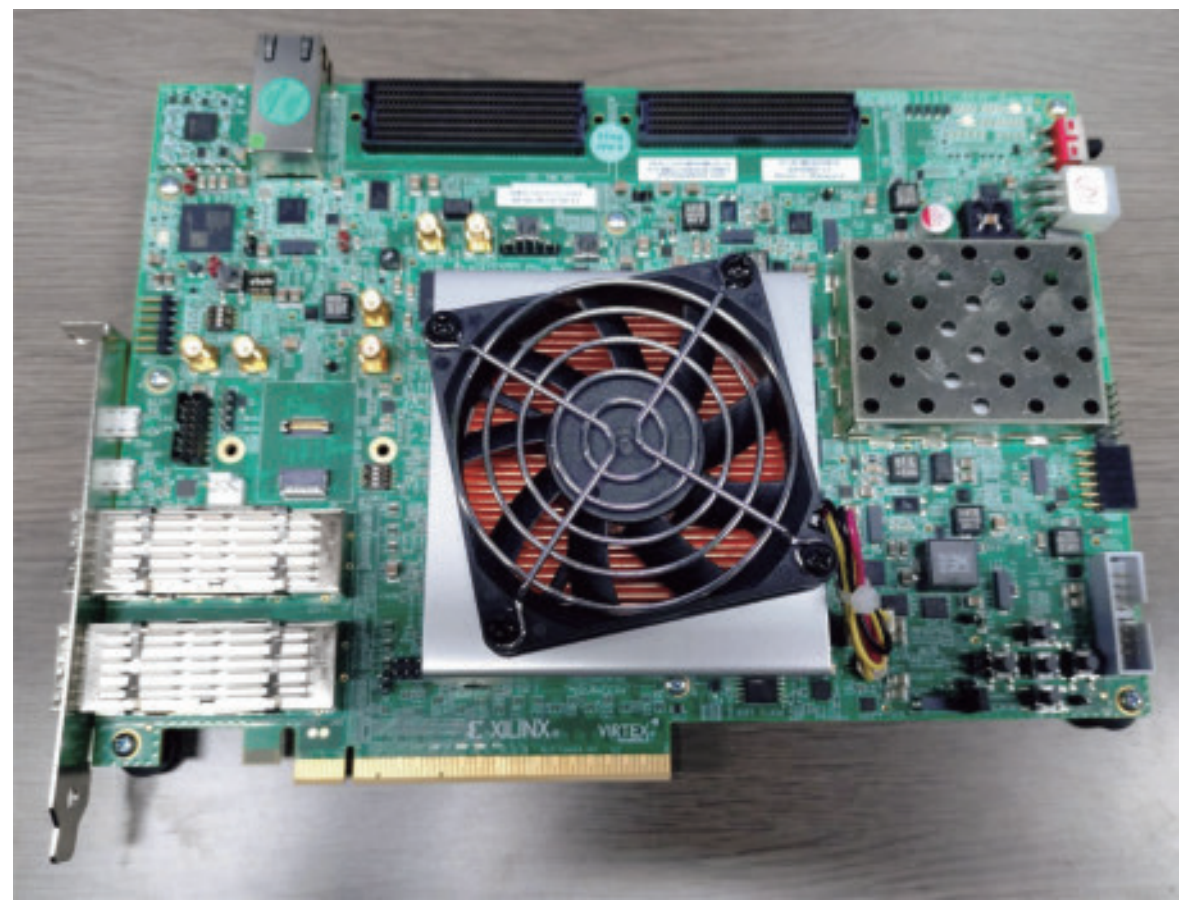
Fig. 3-18. External view of the FPGA.

### 3-3-d. Benchtop Holographic Display Device

HMDs and HMPs require users to wear devices on their heads, which can cause issues such as device weight and discomfort during use. To address these issues, this study investigates a benchtop holographic display device that places no physical obstructions near the viewpoint and provides a comfortable, non-intrusive viewing experience.

#### Wide-Field-of-View Display Device

In general, holographic display devices suffer from a limited field of view. The benchtop display device developed in this study achieves a wide viewing zone, realizing a field of view of 12.8 degrees [29]. In this system, in addition to a 4f optical system, two additional lenses are used to magnify the reconstructed image. Fig. 3-19 shows the external appearance of the device, and Fig. 3-20 presents the reconstructed image. The reconstructed image can be observed through a large-aperture lens.

Furthermore, to enable binocular viewing, a display device employing independent optical systems for the left and right eyes was also developed. However, because the interpupillary distance varies among users, conventional approaches required physical adjustment of the spacing between the two devices. In this study, an alternative approach was adopted in which the reconstruction illumination light is electronically switched to modify the viewing zone. The pupil position is detected using an eye tracker, allowing flexible adaptation to individual differences in interpupillary distance as well as small variations in face position.

#### Wide Viewing-Zone Display Device

Fig. 3-22 shows a display device with an expanded viewing zone, achieving a viewing angle of 32 degrees, which allows observation from a wide range of positions [2, 17, 33, 35]. This capability enables

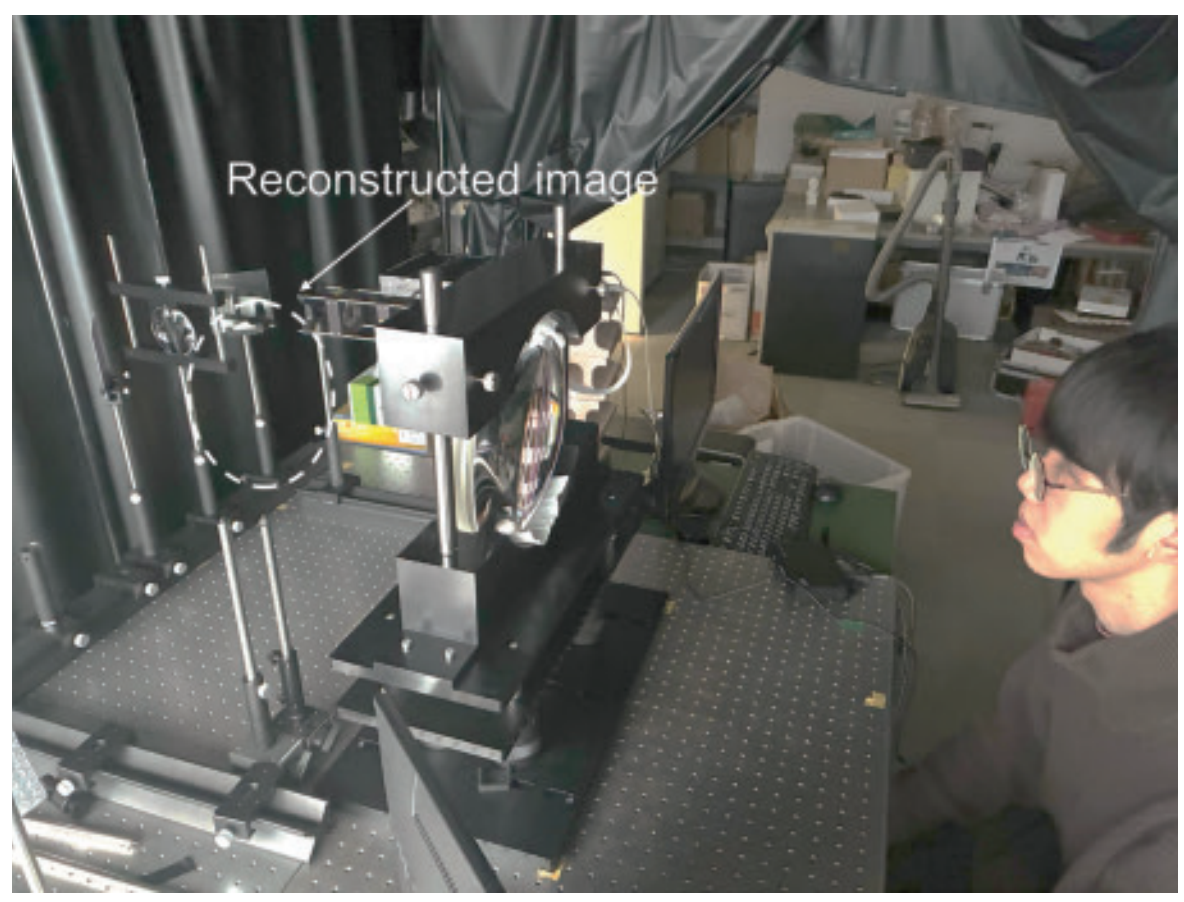

Fig. 3-19. External view of the wide FOV holographic display device.

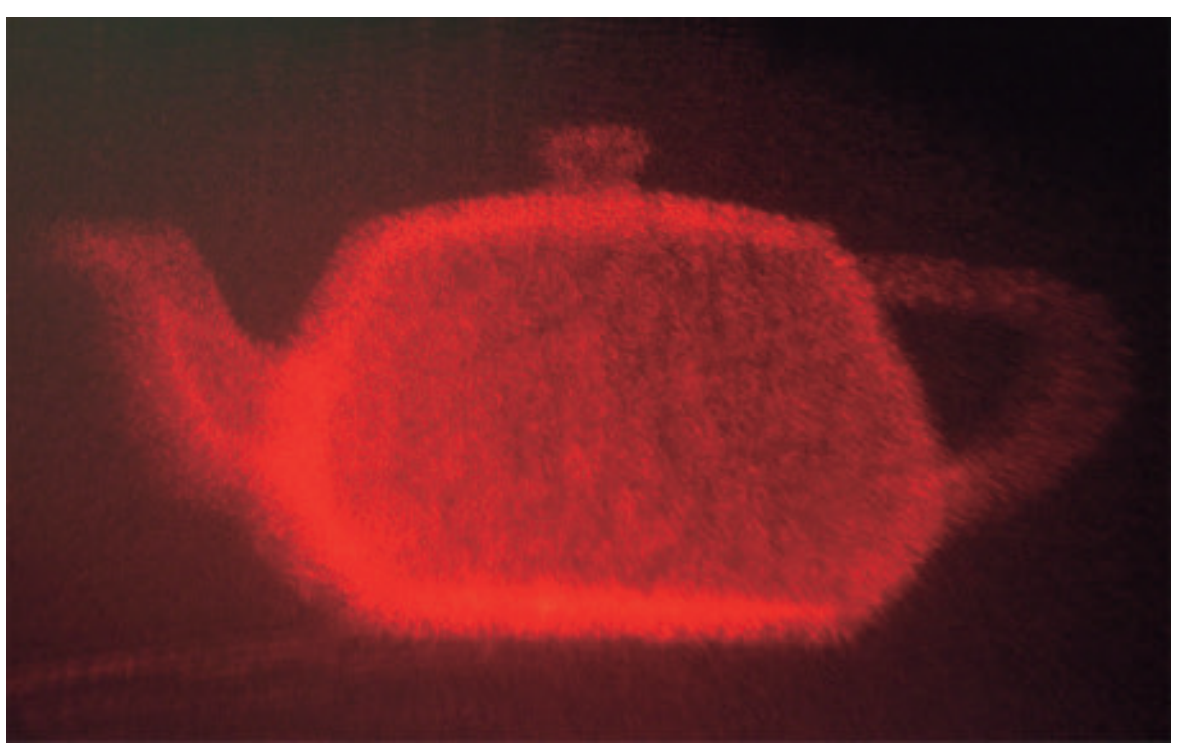
Fig. 3-20. Reconstructed image observed with the wide FOV holographic display device.

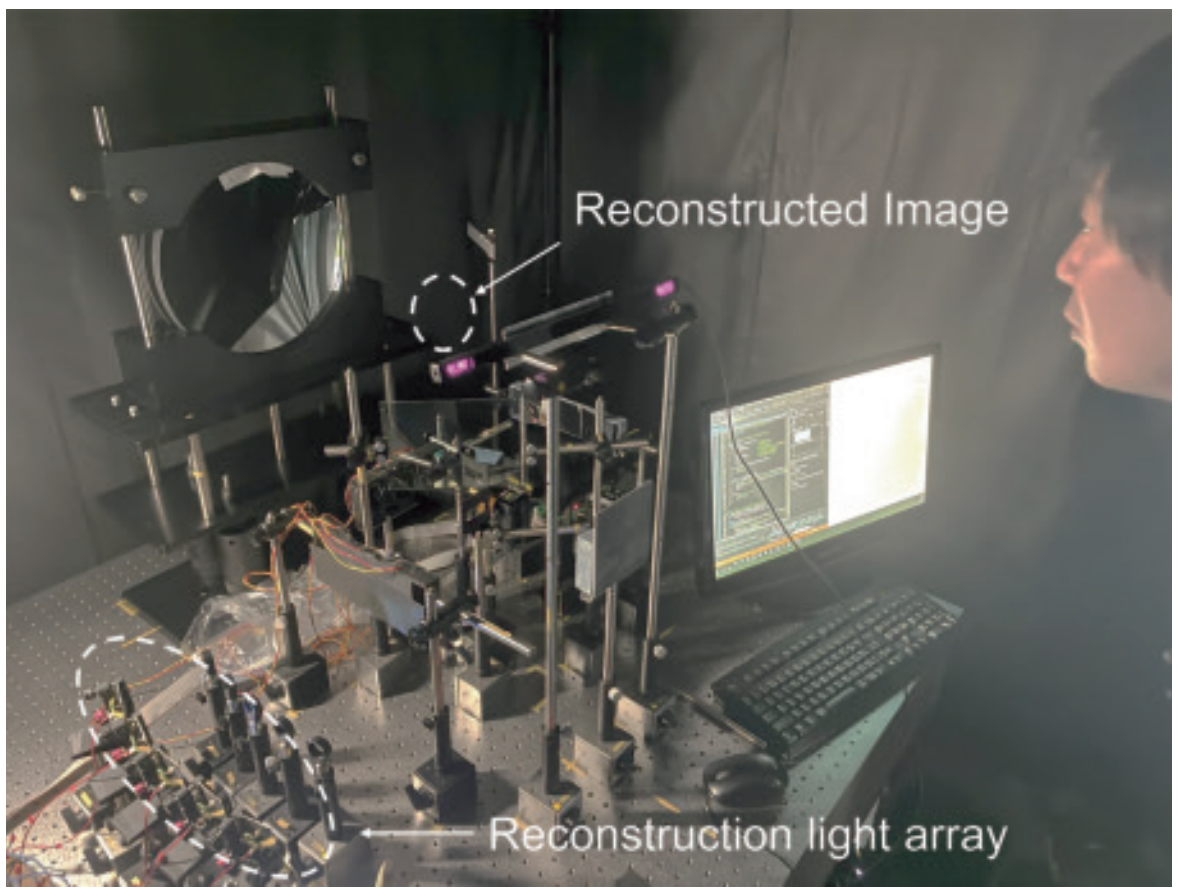

Fig. 3-21. Configuration of the wide viewing-zone holographic display device.

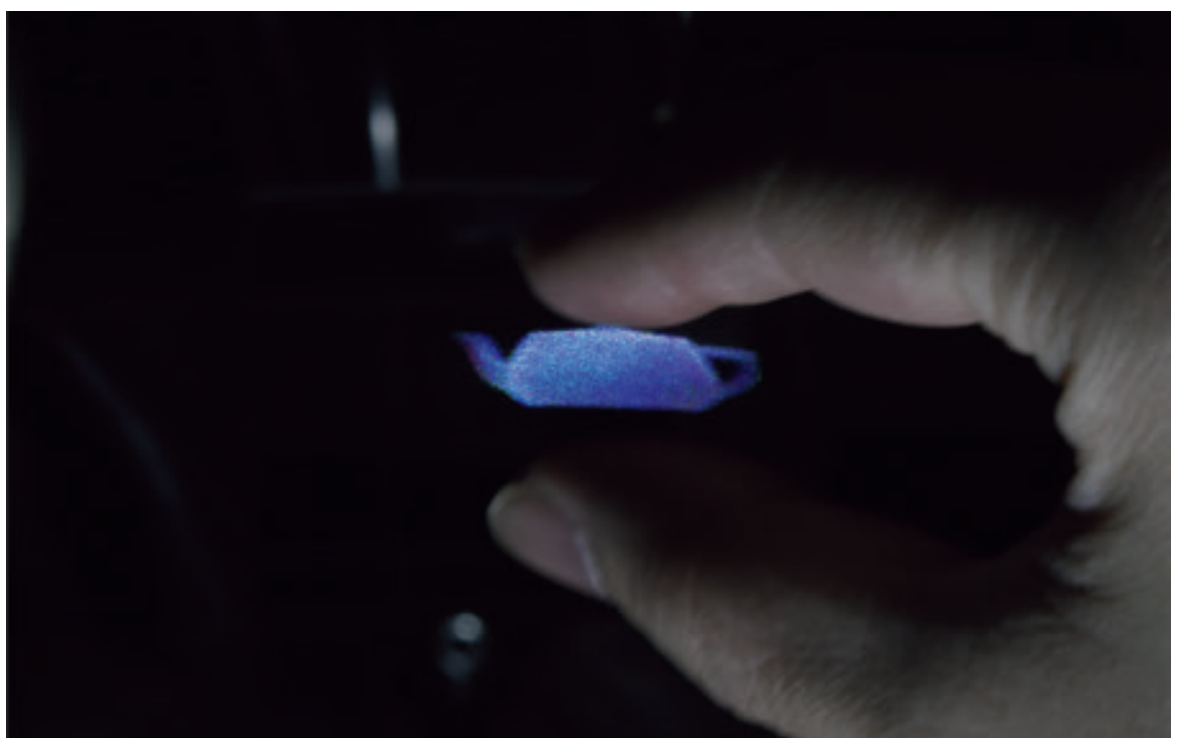
Fig. 3-22. Reconstructed image appearing to float in space.

reconstructed images to be observed from various directions, providing three-dimensional imagery with pronounced motion parallax and occlusion effects. In addition, the reconstructed image appears to float in midair at a position where it can be touched by a finger, as shown in Fig. 3-21, indicating the potential for future interaction using direct finger input.

In this system, the viewing zone is expanded by detecting the viewer's gaze position using an eye tracker and electronically switching the reconstruction illumination light. A notable feature of this approach is that it is realized using only a single spatial light modulator (SLM), which is typically a high-cost component. This design significantly reduces both system cost and complexity while maintaining a wide viewing zone.

### 3-3-e. Reduction of Speckle Noise

Speckle noise is a characteristic type of noise in holographic displays and appears as granular noise in reconstructed images. For the practical realization of holographic display devices, it is essential to reduce this noise. Various approaches have been proposed, such as using specialized spatial light modulators (SLMs) or generating hologram data that suppress speckle noise. In this study, speckle reduction was investigated using a method that modifies only the reconstruction illumination light, without changing the hologram data or the SLM.

#### Reduction of Temporal Coherence Using Yellow Phosphor Excitation and a Frequency Filter

We investigated a method for reducing the temporal coherence of the reconstruction illumination light [1]. In this approach, a white light source is generated using yellow phosphor excitation, and the spectral bandwidth is controlled using a frequency filter. As the spectral bandwidth increases, the temporal coherence decreases, leading to a reduction in speckle noise. Experimental results demonstrated that speckle noise was reduced by approximately 30%. In addition, compared with light sources having low temporal coherence such as light-emitting diodes, the proposed light source provides higher brightness, confirming its effectiveness as an illumination source for holographic display devices.

#### Reduction of Spatial Coherence by Fiber Vibration

We investigated a method for reducing the spatial coherence of the light source, a technique based on vibrating a multimode optical fiber [3]. This method achieved a 43% reduction in speckle noise. However, due to the large numerical aperture of the multimode fiber, some degradation in image resolution was observed.

## 3-4. Hologram Data Generation Using the Human Visual System

Hologram data involve a large amount of information, and a hologram size of 4K requires a communication bandwidth of 6 Gbps. In practical applications, even larger data volumes are expected, making data compression indispensable for transmission. In addition, as described above, there is also the issue that generating hologram data requires a significant amount of computation time.

To address this issue, this study investigates a compression method that exploits characteristics of the human visual system. Humans have a visual field of approximately 160° horizontally and 135° vertically; however, regions with high visual acuity are limited to a narrow area of approximately 3° corresponding to the fovea. Focusing on this characteristic, foveated rendering has been proposed in the fields of VR and AR as a high-speed rendering technique. As shown in **Fig. 3-23**, foveated rendering achieves perceptually high-resolution rendering by rendering the region corresponding to the observer's gaze center at high resolution while rendering the surrounding regions at lower resolution.

In this study, we apply this concept by using high-resolution hologram data for the foveal region and lower-resolution hologram data for the peripheral regions. The advantages of this approach include improved compression efficiency of hologram data and a reduction in computation time.

Accordingly, we developed an algorithm for generating hologram data that display reconstructed images with reduced resolution in peripheral regions. In this algorithm, the resolution of the reconstructed image is reduced by limiting the computation region of the wavefront data emitted from each point light source and recorded on the hologram plane to a restricted area referred to as a *sub-hologram*. Here, when the size of the sub-hologram is denoted as $N$ [pixels], the corresponding angular resolutions are as follows:

- For $N$=250, the angular resolution is 0.0564 deg,
- For $N$=500, the angular resolution is 0.0282 deg,
- For $N$=1000, the angular resolution is 0.0141 deg,
- For $N$=2000, the angular resolution is 0.0071 deg.

### 3-4-a. Subjective Evaluation of Reconstructed Images for Peripheral Vision

In this study, a subjective evaluation experiment was conducted with 23 participants in order to realize foveated rendering for computer-generated holography (CGH) using a general-purpose display device and a simple computation method that takes human visual characteristics into account. Although various types of optical systems can be used for CGH reconstruction, this study applies foveated rendering to a Fresnel-type optical system [37, 38].

The field of view of the optical system used in this experiment was 15.8° × 7.9°, which is sufficiently large compared with the 3° field of view corresponding to the human fovea. In addition, the maximum angular resolution was 0.0037° × 0.0074°, which is sufficiently high such that the resolution limit does not affect subjective evaluation.

In this experiment, two parameters of foveated rendering were investigated. The first parameter was the size of the foveated region, and the second parameter was the number of pixels in the peripheral region. The experiment compared reconstructed images rendered entirely at the highest resolution with reconstructed images in which foveated rendering was applied at various ratios. The Wilcoxon signed-rank test was used for statistical evaluation.

The stimuli presented to the participants consisted of three object images. In all cases, the objects were positioned at a distance of 1 m from the viewpoint. One torus-shaped object was placed at the center, and two star-shaped objects were placed symmetrically on the left and right sides. The torus-shaped object was rendered at the highest resolution (FULL) in all conditions. The positions and pixel numbers of the star-shaped objects were varied depending on the experimental conditions.

The star-shaped images (B) were positioned at angular separations of 3, 5, 7, and 9 degrees from the torus-shaped image (A). The number of pixels was varied by changing the sub-hologram size to $N$=250, 500, 1000, and 2000 pixels.

In this evaluation experiment, a total of 16 combinations were tested for the reference hologram B relative to the reconstructed image A rendered entirely at the highest resolution (angular resolution of 0.0037°). These combinations consisted of four foveated region sizes 3, 5, 7, and 9 degrees and four pixel numbers $N$=250, 500, 1000, and 2000.

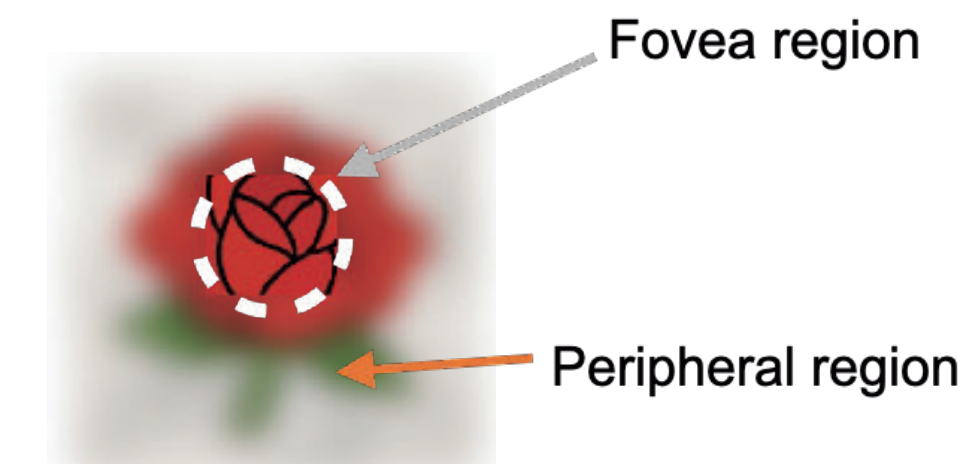

Fig. 3-23. Concept of foveated rendering based on human visual characteristics.

The reconstructed image rendered entirely at the highest resolution is referred to as A, while the reconstructed image with foveated rendering applied is referred to as B. Participants were asked the question, "*How did image B compare with image A?*"

In the experiment, not only images A and B were presented, but a stimulus-reset period was also introduced by presenting only the torus-shaped object at three stages: before the presentation of A, during the transition from A to B, and after the presentation of B. The detailed procedure of the stimulus presentation experiment is shown in Table 1.

Specifically, the experiment began with the presentation of only the torus-shaped object, followed by the presentation of reconstructed image A. Participants were instructed to memorize image A. Subsequently, only the torus-shaped object was presented again, and after a verbal countdown of "3, 2, 1," reconstructed image B was presented. Participants observed image B for 5 seconds. When only the torus-shaped object was presented again, the observation phase ended, and participants responded to the questionnaire.

The evaluation was conducted using a five-point subjective scale to assess how the perceived resolution of image B changed relative to the reference image A:
1 (1): worsened (changed),
2 (2): slightly worsened (slightly changed),
3 (3): no change,
4 (2): slightly improved (slightly changed),
5 (1): improved (changed).

This evaluation experiment was repeated for all 16 combinations of the 4 × 4 parameter settings. The order of stimulus presentation was randomized. Afterward, for statistical analysis of the participants' evaluation scores and the test results, the five-point ratings were converted into a three-level scale, as shown in Fig. 3-24. The correspondence of the evaluation scores is indicated in parentheses.

### 3-4-b. Acceleration of Hologram Data Computation

In this algorithm, the computation region per point light source is restricted by the use of sub-holograms. As a result, reducing the resolution directly decreases the computational load, enabling faster computation. As shown in Fig. 3-25, a speedup of more than eight times was achieved compared with high-resolution computation.

In practice, computation for the foveated region is also required; however, since this region spans only approximately 5 degrees, most of the hologram area is computed at low resolution. Consequently, even when including the computation of high-resolution hologram data for the foveated region, the overall hologram generation process remains significantly faster.

### 3-4-c. Achievement of a High Compression Ratio

Because the amount of information contained in hologram data is reduced by the use of sub-holograms, a higher compression ratio can be achieved compared with conventional hologram data [14, 32]. Fig. 3-26 shows the relationship between the file size compressed using a conventional lossless compression method and the sub-hologram size (resolution). The results confirm that by reducing the resolution, the compression ratio can be improved by more than a factor of three.

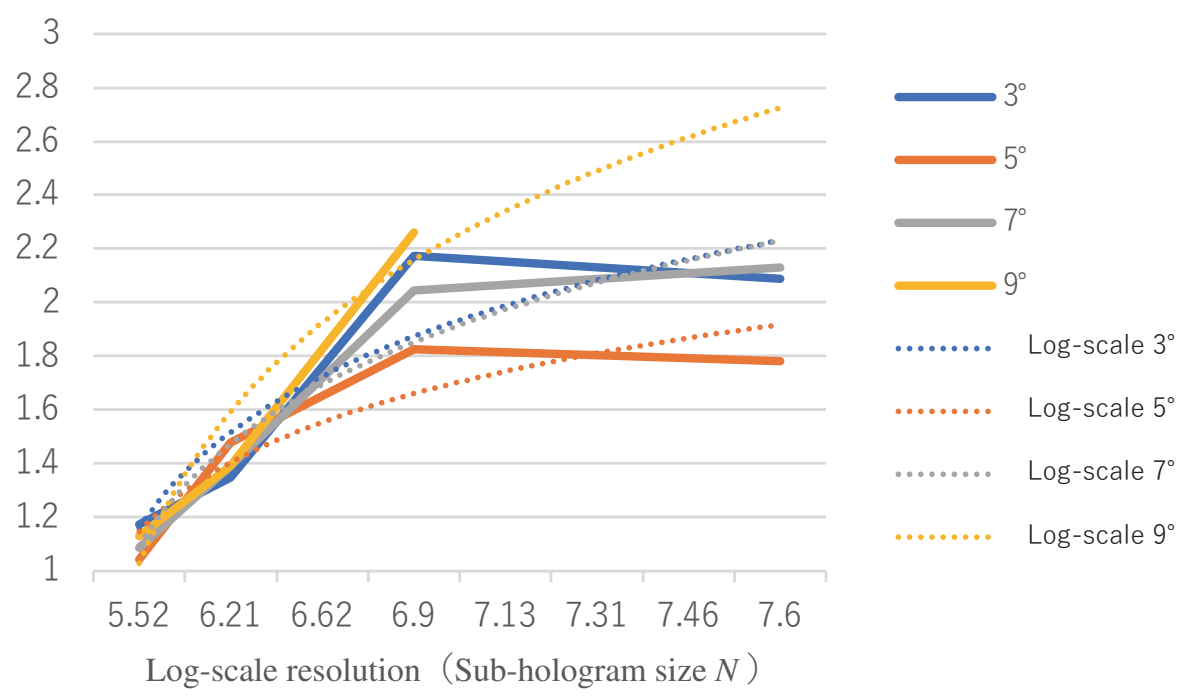

**Fig. 3-24.** Mean values of the modified three-level subjective evaluation scores and their logarithmic representation.

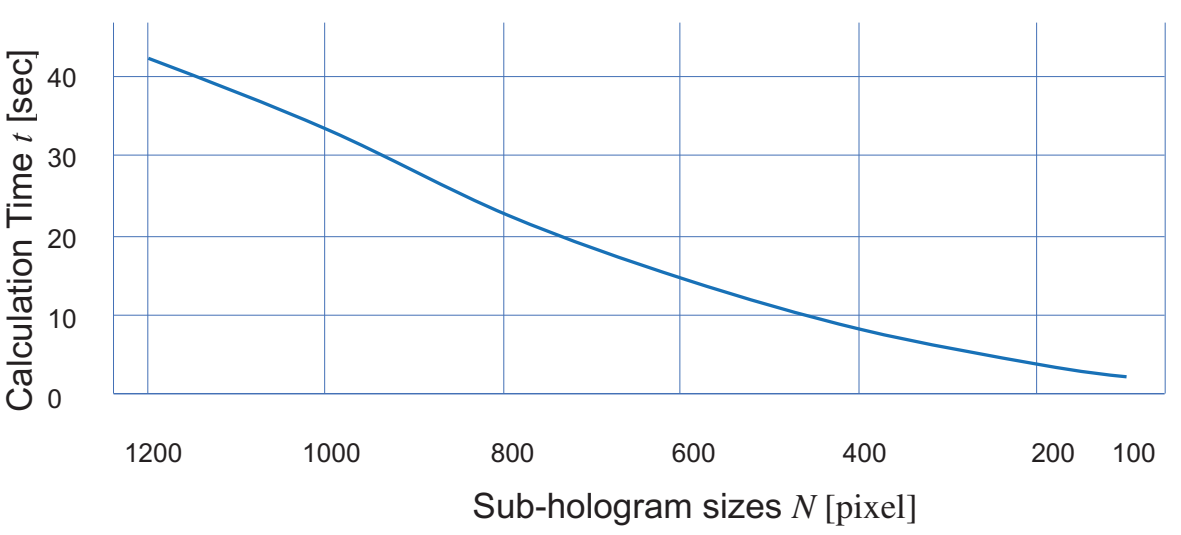

**Fig. 3-25.** Computation time comparison between high-resolution hologram generation and sub-hologram-based computation.

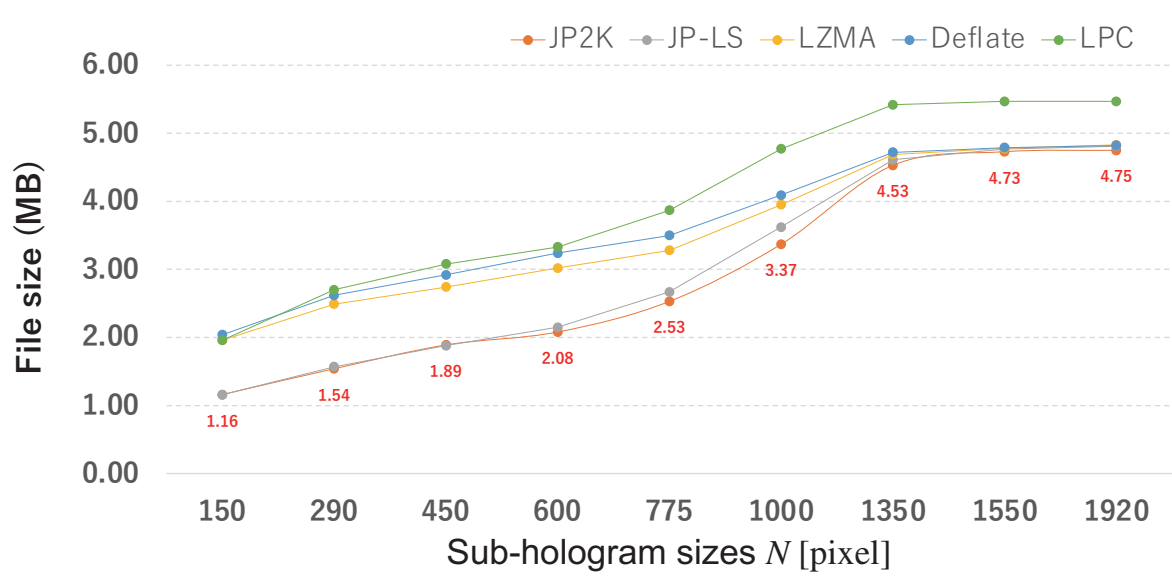

**Fig. 3-26.** Compressed file size versus sub-hologram resolution compared with conventional hologram data.

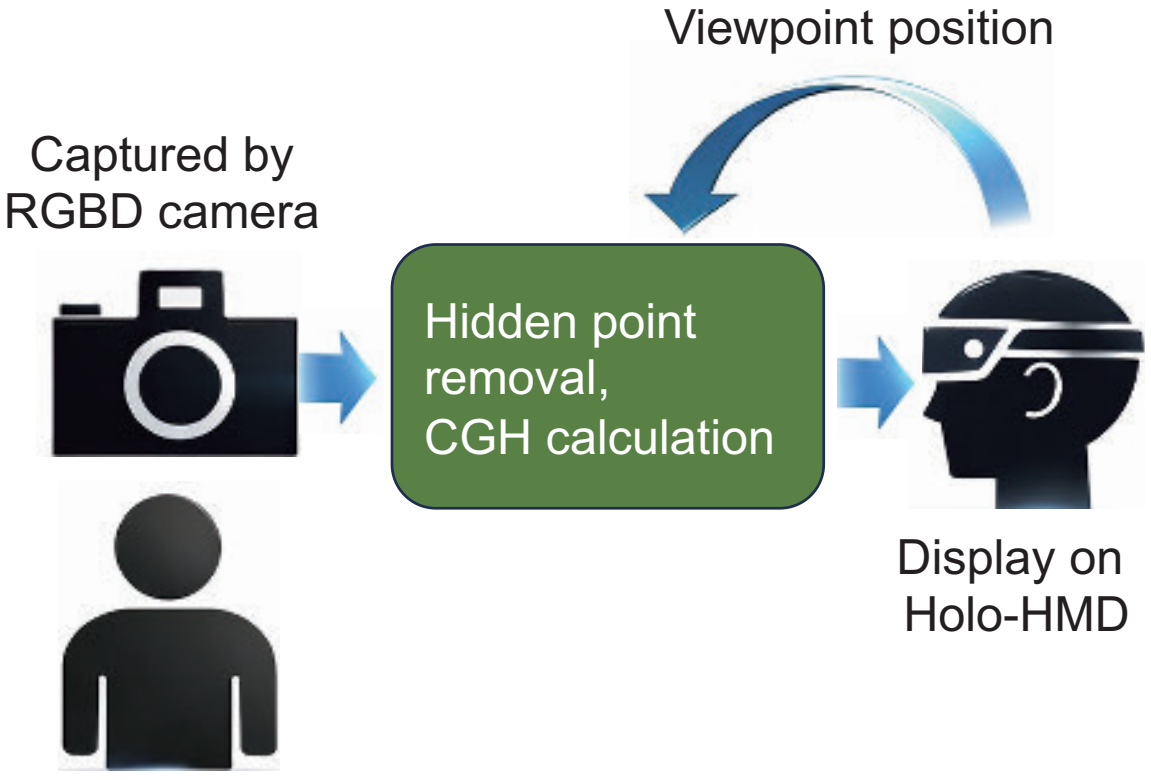

**Fig. 3-27.** System configuration for the free-viewpoint experiment using real-world data.

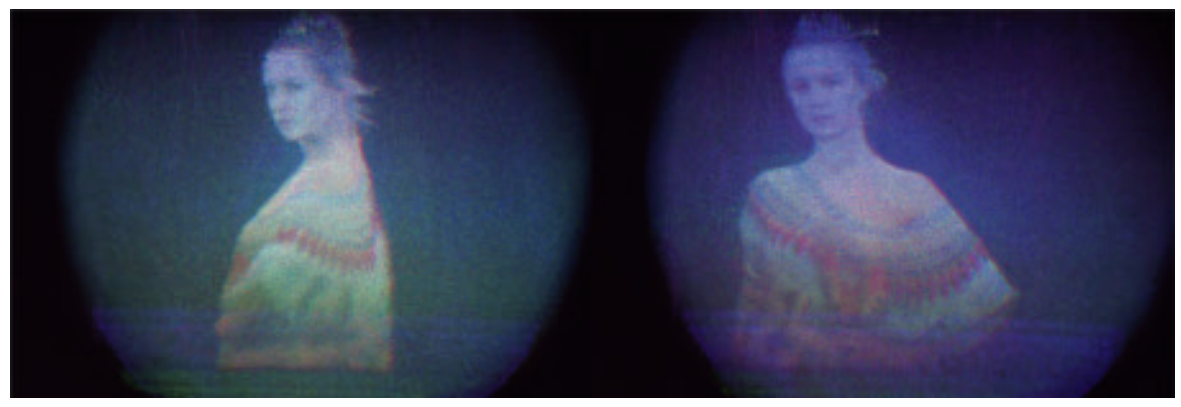

**Fig. 3-28.** Reconstructed images observed from surrounding viewpoints using real-world captured data.

## 3-5. Free-Viewpoint Experiment Using Real-World Data

As a system test, display experiments were conducted in which 3D data captured from real-world objects were displayed on a Holo-HMD while changing the viewpoint [26, 27]. A block diagram of the system configuration is shown in Fig. 3-27. The 3D data were captured using *4DVIEWS* provided by Crescent Inc., a collaborative research organization in this project, and were obtained as polygon data and point cloud data.

Based on the head rotation information acquired from the motion sensor of the Holo-HMD, viewpoint-dependent object rotation and hidden surface removal were applied to the acquired point cloud data. As a result, reconstructed images that follow the observer's head motion were successfully presented.

Fig. 3-28 shows reconstructed images observed from various surrounding viewpoints using the captured data of a female subject. Depending on the change in viewpoint position, occlusion and motion parallax effects were observed, enabling visualization of the object even from its lateral directions.

From these results, it was confirmed that free-viewpoint display supporting six degrees of freedom (6DoF) can be achieved by holographic display even when using real-world captured data.

# 4. Conclusion

Holo-TV has long been expected to be realized as a three-dimensional television system. However, its practical realization has been considered a long-term challenge due to the many technical issues that must be resolved and the high level of difficulty involved. The Holographic Viewing System (HVS) proposed in this study is an effective approach for broadcast-type holographic communication. It enables multiple users to independently view three-dimensional images from their own free viewpoints and is also capable of supporting different types of holographic display devices.

In this study, we investigated the individual elemental technologies required to realize HVS and successfully established effective fundamental technologies for broadcast-type holographic communication. These results demonstrate the feasibility of holographic communication based on the proposed HVS.

On the other hand, this study mainly focused on the investigation of individual elemental technologies. In future work, demonstration experiments using an end-to-end system that integrates these technologies will be required. Furthermore, continued efforts are necessary to achieve further acceleration of computation and improvement of data compression efficiency.

# 5. Social Impact and Practical Applications

The outcomes of this research provide a new means of information presentation that enables more natural sharing of three-dimensional information. In the fields of education and healthcare, the proposed technology is expected to be applied to remote learning materials and diagnostic support systems that allow intuitive understanding of three-dimensional structures. In the manufacturing industry, it is anticipated to contribute to improved work efficiency by clearly visualizing design information and operational procedures.

Furthermore, this research serves as an example of services premised on future ultra-high-capacity and low-latency communication networks, such as Beyond 5G. As such, it is expected to contribute to the effective utilization of next-generation communication infrastructures and to studies toward their social implementation.

# 6. Project History

- January 2023: Project started.
  Initial research group consisted of Yuji Sakamoto (principal investigator at Hokkaido University), Takahiro Yamanoi, and Kang Seog (all from Hokkaido University).
- December 2024: Following the retirement of Yuji Sakamoto, Seok Kang became the principal investigator at Hokkaido University.
- April 2025: Yoshinori Dobashi, Manabu Yamamoto, and Takashi Hikage (all from Hokkaido University) joined the project as researchers.
- March 2026: Project scheduled to be completed.

## Acknowledgments

These research results were obtained from the commissioned research (No. PJ012368C06801) by National Institute of Information and Communications Technology (NICT), Japan.

## Research achievements

### Patent

Title of the invention: Hologram Computing Apparatus, Hologram Computing Method, and Hologram Image Display System
Application number: JP2024-190104
Filing date: October 29, 2024

### Exhibition

Presentation destination: IDW'24 University Exhibition
Title: Holometric video streaming
Date: December 4th to December 6th, 2024
Place: Sapporo Convention Center
Content: Demonstration with holographic head mounted display (Holo HMD)

### Press coverage

Presentation destination: HBC (Hokkaido Broadcasting Co., Ltd.) Web Magazine "Sitakke"
Title: Is the SF world a reality? Hokkaido University is trying to develop the ultimate 3D imaging technology「Electronic Holography」!
Announcement date: March 26, 2024
URL: https://sitakke.jp/post/9882/

### Award

Host organization: OPTICA
Award title: The 2024 Best Paper Prize for Applied Optics
Date: October 26, 2025

## Academic papers and conference presentations

**FY2023 (April 2023 ~ March 2024)**

[Journal paper]

[1] T. Yoneyama and Y. Sakamoto, "Speckle control for electro-holographic display using high-brightness yellow phosphor light source in projector," Opt. Eng., vol. 62, no. 8, p. 083103, Aug. 2023.

[2] R. Watanabe and Y. Sakamoto, "Enlargement of the viewing zone and size of a reconstructed image in electro-holography using multiple reconstruction lights by eye-tracking," Appl. Opt., vol. 63, no. 7, pp. B76-B84, Mar. 2024.

[International Conference]

[3] K. Matsuno, Y. Ito, and Y. Sakamoto, "Controlling Speckle and Resolution of Reconstructed Images by Vibrating Multimode Optical Fiber with a Cylindrical Piezoelectric Transducer in Electro-Holography," Proc. IDW '23, p. 711, Dec. 2023.

[4] T. Sakakihara and Y. Sakamoto, "3DoF holographic head-mounted display with motion sensor based on fast calculation algorithm for CGH," Proc. SPIE, vol. 12910, pp. 126-131, Jan. 2024.

[5] N. Kikuchi and Y. Sakamoto, "Proposal of head-mounted projector for field of view enlargement in electro-holography," Proc. SPIE, vol. 12910, p. 129100L, Jan. 2024.

[Domestic Conference]

[6] K. Yamauchi, A. Kashiwagi, and Y. Sakamoto, "Distributed Calculation of Light Propagation for Computer-generated Hologram using Cloud Cluster," ITE Technical Report, vol. 47, pp. 29-32, Sep. 2023.

[7] S. Iinuma and Y. Sakamoto, "Acceleration of Computer-Generated Hologram for Head-Mounted Display Using FPGA," 2023 Joint Convention, the Hokkaido Chapters of the Institutes of Electrical and Information Engineers, pp. 163-164, Oct. 2023.

[8] K. Ono and Y. Sakamoto, "Comparison of Computation Time for Computer Generated Hologram with Different Programmable Memories in CUDA," 2023 Joint Convention, the Hokkaido Chapters of the Institutes of Electrical and Information Engineers, pp. 161-162, Oct. 2023.

[9] T. Sakakihara and Y. Sakamoto, "Effort to Develop a Practical Head- Mounted Display Using Electro-Holography," 2023 Joint Convention, the Hokkaido Chapters of the Institutes of Electrical and Information Engineers, pp. 144-145, Oct. 2023.

[10] K. Yamauchi, A. Kashiwagi, and Y. Sakamoto, "Study on Generating Photorealistic Holographic Movies using Cloud Cluster," ITE Technical Report, vol. 47, no. 38, pp. 37-40, Dec. 2023.

[11] K. Yamauchi, A. Kashiwagi, and Y. Sakamoto, "Holographic Video Game with Photorealistic Rendering using Cloud Cluster," ITE Technical Report, vol. 48, no. 1, pp. 45-48, Jan. 2024.

[12] S. Iinuma and Y. Sakamoto, "Verification of hologram calculation circuit for head-mounted display using FPGA," HODIC Student Symposium 2024, pp. 1-4, Feb. 2024.

[13] K. Ono and Y. Sakamoto, "Implementation and its acceleration of mirror image reflected on Bézier surfaces in computer-generated hologram," HODIC Student Symposium 2024, pp. 5-8, Feb. 2024.

[14] D. Kim and Y. Sakamoto, "Lossless compression review of the sub- hologram," HODIC Student Symposium 2024, pp. 47-50, Feb. 2024.

**FY2024 (April 2024 ~ March 2025)**

[Journal paper]

[15] R. Kon and Y. Sakamoto, "Real-time generation of computer-generated hologram for 360-degree panoramic views using cylindrical object light," Opt. Eng., vol. 63, no. 6, p. 065101, June 2024.

[16] H. Arai, K. Ono, and Y. Sakamoto, "CGH calculation algorithm for expressing reflection on a curved mirror surface," Opt. Express, vol. 32, no. 21, pp. 36469–36488, Oct. 2024.

[17] R. Watanabe, S. Kang, and Y. Sakamoto, "Binocular holographic display with a wide viewing zone using eye-tracking and multiple reconstruction lights," Appl. Opt., vol. 64, no. 4, pp. 984–994, Feb. 2025.

[18] A. Kashiwagi, K. Ono, and Y. Sakamoto, "Fast calculation method for CCGHs applicable to various planar computer-generated hologram computation methods," Appl. Opt., vol. 64, no. 7, pp. B38–B48, Mar. 2025.

[19] N. Kikuchi, S. Kang, and Y. Sakamoto, "Head-mounted projector with a wide visual-field and three degrees of freedom in electro-holography," Appl. Opt., vol. 64, no. 7, pp. B58–B64, Mar. 2025.

[International Conference]

[20] T. Sakakihara and Y. Sakamoto, "Holographic Head-mounted Display Capable of Displaying AR in accordance with User's Attitude Angle," DH 2024, p. W4A.4, June 2024.

[21] S. Iinuma and Y. Sakamoto, "Design of calculation circuit of computer-generated hologram for holographic head-mounted display," IMID 2024, p. 600, Aug. 2024.

[22] K. Ono, H. Arai, and Y. Sakamoto, "Generation of Mirror Images of Bézier Surfaces in Computer-generated Hologram," IMID 2024, p. 594, Aug. 2024.

[23] N. Kikuchi and Y. Sakamoto, "Head-Mounted Projector for Enlarging Field-of-View Corresponding to User's Attitude Angle in Electronic Holography," IMID 2024, p. 681, Aug. 2024.

[24] T. Sakakihara and Y. Sakamoto, "Holographic Head-mounted Display with Real-time Calculation using Fast Calculation Algorithm for Computer-generated Hologram," IWH 2024, Paper S3-3, Dec. 2024.

[25] K. Ono, S. Kang, and Y. Sakamoto, "Fast calculation method for curved surface mirror reflections using subdivision in computer-generated hologram," Proc. SPIE, vol. 13390, p. 133900L, Jan. 2025.

[26] S. Nishita et al., "Display method of real-world data captured by an RGBD camera on a holographic head-mounted display reacting to viewpoint movement," Proc. SPIE, vol. 13390, p. 133900M, Jan. 2025.

[Domestic Conference]

[27] S. Nishita, T. Sakakihara, and Y. Sakamoto, "Generating CGH using real data and playing it on a Holographic head-mounted display," 2024 Joint Convention, the Hokkaido Chapters of the Institutes of Electrical and Information Engineers, pp. 99–100, Oct. 2024.

[28] A. Umenai, A. Kashiwagi, and Y. Sakamoto, "Method for Object Light Conversion Algorithm for Various Electro-holography Display," HODIC Student Symposium 2025, pp. 5–8, Mar. 2025.

[29] T. Ogawa and Y. Sakamoto, "Proposal of a Method for Enlarging and Full- Colorizing the Reproduced Image in Extended-View Optics," HODIC Student Symposium 2025, pp. 1–4, Mar. 2025.

[30] T. Miyahara and Y. Sakamoto, "Generation of Holographic Video in High- Speed Calculation System with GPU Cluster and Physical Simulation," HODIC Student Symposium 2025, pp. 25–28, Mar. 2025.

**FY2025 (April 2025 ~ March 2026)**

[Journal paper]

[31] K. Ono et al., "Fast computer-generated hologram calculation algorithm for mirror images reflected on mirror surface of Bézier surfaces using subdivision," Opt. Eng., vol. 64, no. 7, p. 075102, July 2025.

[32] D. Kim et al., "Lossless compression of CGH via resolution-controlled sub-hologram," Appl. Opt., vol. 64, no. 36, pp. 10678–10690, Dec. 2025.

[International Conference]

[33] S. Iwai et al., "Enlargement of Viewing Zone by Multiple Light Sources in Holographic Display with Wide Field-of-View," IMID 2025, p. 713, Aug. 2025.

[34] A. Umenai et al., "Object Light Conversion Algorithm for Various Display Devices of Electro-holography using Backpropagation Calculation," IMID 2025, p. 530, Aug. 2025.

[35] T. Ogawa et al., "Enlargement of Reconstructed Image and Full-colorization in Holographic Display with Wide Viewing-zone," IMID 2025, p. 532, Aug. 2025.

[36] T. Miyahara et al., "Implementation of Holographic Video Game in High-Speed Generation System with Physical Simulation and GPU Cluster," IMID 2025, p. 531, Aug. 2025.

[37] T. Yamanoi et al., "Sensory Evaluation of Human Vision for Foveated Rendering of Computer-Generated Holograms," ISIS 2025, pp. 526–531, Nov. 2025.

[Domestic Conference]

[38] T. Yamanoi, A. Muramoto, R. Yoshioka, Y. Sakamoto, and S. Kang, "Sensory Evaluation of Foveated Rendering in Computer-Generated Holograms," FSS2025, pp. 668–671, Sep. 2025.

[39] S. Iinuma, S. Kang, and Y. Sakamoto, "Design of Parallel High-speed Computing Circuits for Realizing Holometric Video Streaming," ITE Winter Annual Convention 2025, 21C–1, Dec. 2025.

[Invited lecture]

[40] Y. Sakamoto, "Hologram Data Generation Method for Various Electro-Holography Displays from Common Object Light," IWH 2025, Paper WII–I–04, Nov. 2025.

[41] Y. Sakamoto, "Holometric Video Streaming for Multiple Users to Simultaneously Experience Free Viewpoints," IDW '25, Paper 3D2–1, Dec. 2025.

[Publicly available scholarly manuscripts]

[42] KODAI ONO et al., "A High-Speed CGH Calculation Method for Mirror Images on Bézier Surfaces using Optical Path Length Minimization," arXiv:2601.06459 [physics.optics], 2026.

[43] Taichi Sakakihara et al., "Head-wearable Holographic Head-mounted Display with 6 Degrees of Freedom," arXiv:2601.15581 [physics.optics], 2026.

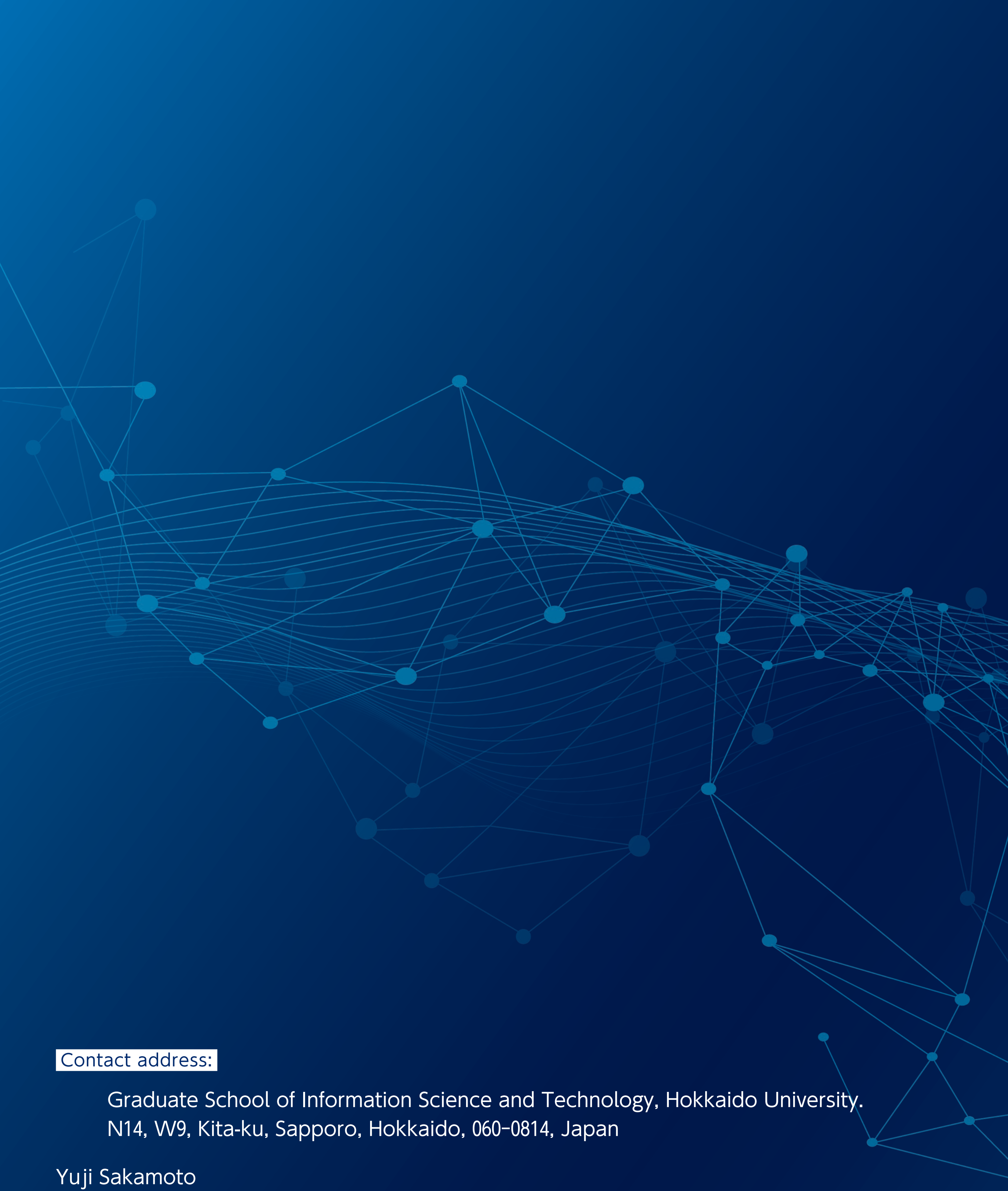

Contact address:

Graduate School of Information Science and Technology, Hokkaido University.
N14, W9, Kita-ku, Sapporo, Hokkaido, 060-0814, Japan

Yuji Sakamoto
hologram.resercher.yuji@gmail.com

Seok Kang
kssrh@ist.hokudai.ac.jp